\documentclass[12pt,prd,preprintnumbers,eqsecnum,nofootinbib]{revtex4}  
\usepackage{epsfig}
\usepackage{amsmath}
\usepackage{amssymb}
\usepackage{latexsym}
\usepackage{multirow}
\usepackage{times}

\begin{document}
\preprint{CERN-TH/2002-202}
\preprint{KUL-TF-02/20}

\title{\vspace*{30pt}
\LARGE\bf Neutrino masses from universal Fermion
mixing\vspace{24pt}}

\author{\large R.\ Gastmans}

\affiliation{Institute for Theoretical Physics, Katholieke Universiteit
Leuven, B-3001 Leuven, Belgium\vspace{10pt}} 

\author{\large Per Osland}

\affiliation{Department of Physics, University of Bergen, All\'{e}gaten 55,
N-5007 Bergen, Norway\vspace{10pt}}

\author{\large Tai Tsun Wu}

\affiliation{Gordon McKay Laboratory, Harvard University, Cambridge,
Massachusetts 02138, U.S.A., and\\ 
Theoretical Physics Division, CERN, CH-1211 Geneva 23,
Switzerland\vspace{24pt}} 

\begin{abstract}

If three right-handed neutrinos are added to the Standard Model, then, for the
three known generations, there are six quarks and six leptons. It is then
natural to assume that the symmetry considerations that have been applied to
the quark matrices are also valid for the lepton mass matrices. Under this
assumption, the solar and atmospheric neutrino data can be used to determine
the individual neutrino masses. Using the $\chi^2$ fit, it is found that the
mass of the lightest  neutrino is $(2\text{--}5)\times10^{-3}$~eV, that of the
next heavier neutrino is  $(10\text{--}13)\times10^{-3}$~eV, while the mass of
the heaviest neutrino is $(52\text{--}54)\times10^{-3}$~eV.
\end{abstract}

\maketitle

\newpage
\section{Introduction}\label{sec:I}

In the work of Lehmann, Newton and Wu \cite{Lehmann:1995br}, the
Cabibbo-Kobayashi-Maskawa (CKM) \cite{Cabibbo:yz} matrix is expressed in
terms of the masses of the three generations of quarks:
\begin{equation}
\begin{pmatrix} u \\ d\end{pmatrix}
\begin{pmatrix} c \\ s\end{pmatrix}
\begin{pmatrix} t \\ b\end{pmatrix}.
\end{equation}
This is accomplished by introducing a new horizontal symmetry. Some of the
earlier attempts in this direction are given in
\cite{Harari:1978yi,Fritzsch:1995dj}, while some of the more recent work on
this topic is to be found in \cite{Altarelli:gu,Roberts:2001zy}, for example.

Recent experiments at Super-Kamiokande \cite{Fukuda:1998tw,Fukuda:1998fd}
indicate the presence of neutrino oscillations, which would imply that the 
neutrinos are not all massless.  If it is accepted that the neutrinos  are not
massless, then it is most natural in the Standard Model \cite{Glashow:tr} to
introduce three  right-handed neutrinos in addition to the three known
left-handed ones.  In this way, there are six quarks and six leptons.  For
recent reviews  of neutrino physics, see \cite{review}.  Most of the recent
work on neutrino  masses is focused on the ideas of grand unification and the
see-saw  mechanism (see, e.g. \cite{King:1999mb}).

In this paper, the consequences of a universal quark-lepton mixing are
studied.   While many authors favor Majorana masses for the neutrinos, this
universality is most natural if they instead have Dirac masses.\footnote{As
shown in \cite{Bilenky:1980cx}, neutrino-oscillation experiments cannot
distinguish between massive Majorana and Dirac neutrinos.} In other words, the
method of \cite{Lehmann:1995br} is used to express the lepton CKM matrix and
the neutrino mixing matrix in terms of the masses of the three generations of
leptons:
\begin{equation}
\begin{pmatrix} \nu_e   \\ e  \end{pmatrix}
\begin{pmatrix} \nu_\mu \\ \mu \end{pmatrix}
\begin{pmatrix} \nu_\tau \\ \tau \end{pmatrix}.
\end{equation}
Of course the masses of the three charged leptons are accurately known, leaving
as unknown parameters the masses of the three neutrinos. Thus there are three
parameters to be determined instead of seven, the three masses plus the four in
the lepton CKM matrix.

It is the purpose of this paper to use the data from solar neutrinos
\cite{Fukuda:1998fd,Cleveland:1994er,Abdurashitov:1999bv,%
Hampel:1998xg,Ahmad:2001an,Ahmad:2002jz} and
atmospheric neutrinos \cite{Fukuda:1998tw} to determine the three neutrino
masses separately, not only the differences of their squares. As compared with
earlier work on this model \cite{Osland:2000bh},  mixing in the charged lepton
sector is also taken into account.  

In Secs.~\ref{sec:2} and \ref{sec:3}, we review the case of quarks and apply
the model to the case of leptons. In Sec.~\ref{sec:4}, the rotation matrix is
discussed, and in Sec.~\ref{sec:5} we construct the resulting mixing matrix
for charged-current interactions.  In Sec.~\ref{sec:6} we review the relevant
formulas for atmospheric neutrino propagation, whereas Secs.~\ref{sec:7} and
\ref{sec:8} are devoted to the three-flavor MSW problem and solar
neutrinos. In Sec.~\ref{sec:9} we collect the results of the fits to the data,
and Sec.~\ref{sec:10} contains a discussion. Some properties of the mass
matrix are discussed in Appendix~\ref{app:A}, whereas Appendices \ref{app:B}
and \ref{app:C} contain technical details of the analytical solution of the
MSW equations.

\vfil
\section{Review for the Case of Quarks}\label{sec:2}

For the quark mass matrices, the  result of Ref.~\cite{Lehmann:1995br} is
\begin{eqnarray}
\label{Eq:md-nondiag}
M(d)&=&\left(\begin{matrix}
0    & d(d) & 0 \\
d(d) & c(d) & b(d) \\
0    & b(d) & a(d)
\end{matrix}\right), \\[3mm]
\label{Eq:mu-nondiag}
M(u)&=&\left(\begin{matrix}
0    & id(u) & 0 \\
-id(u) & c(u) & b(u) \\
0    & b(u) & a(u)
\end{matrix}\right),
\end{eqnarray}
with
\begin{equation}
\label{Eq:b-vs-c}
b^2(d)=8c^2(d), \quad b^2(u)=8c^2(u).
\end{equation}

The diagonalization of these mass matrices is achieved by
the orthogonal matrices $R(d)$ and $R(u)$, explicitly\footnote{Note 
misprints in Eqs.~(27) and (35) of Ref.~\cite{Lehmann:1995br},  $i\to-i$.}
\begin{eqnarray}
\label{Eq:diagonal-d}
M(d)&=&R(d)\text{diag}(m_d,-m_s,m_b)R^{\text{T}}(d), \\[3mm]
\label{Eq:diagonal-u}
M(u)&=&\text{diag}(i,1,1)R(u)\text{diag}(m_u,-m_c,m_t)R^{\text{T}}(u)
\text{diag}(-i,1,1).
\end{eqnarray}
In terms of $R(d)$ and $R(u)$, the CKM
mixing matrix \cite{Cabibbo:yz} is written as
\begin{equation}
\label{Eq:VKM-R_def}
V=R^{\text{T}}(u)\text{diag}(-i,1,1) R(d).
\end{equation}

For both the $u$ quarks and the $d$ quarks, the number of independent
parameters in the mass matrix is three.
Hence they can be expressed in terms of the three quark masses.
The relations are
\begin{alignat}{2}
\label{Eq:quarks:m1m2m3}
a+c&=\hphantom{\hbox{$-$}}S_1&\mbox{}=m_3-m_2+m_1,\hbox to .83in{}\nonumber
\\
8c^2+d^2-ac&=-S_2&\mbox{}=m_3m_2-m_3m_1+m_2m_1,\hbox to .2in{} \nonumber \\
ad^2&=-S_3&\mbox{}=m_1m_2m_3,\hbox to 1.23in{}
\end{alignat}
where $m_1$, $m_2$, $m_3$ mean respectively $m_u$, $m_c$, $m_t$
and $m_d$, $m_s$, $m_b$ for the $u$ and $d$ quarks.
Because of the known masses, the inequalities
\begin{equation}
\label{Eq:quark-mass-inequality}
m_1 \le m_2 \le m_3
\end{equation}
are used throughout the analysis of Ref.~\cite{Lehmann:1995br}.

In Ref.~\cite{Lehmann:1995br}, it is found that \cite{Jarlskog:1985ht}
\begin{equation} \label{Eq:quarks-J}
|J_{CP}|\simeq 2.6\times 10^{-5}.
\end{equation}
This is consistent with the experimental value of
$(3.0\pm 1.3)\times 10^{-5}$ \cite{Harrison:1998yr}.
This experimental value is expected to improve significantly
in the near future.

\section{Application to the Case of Leptons}\label{sec:3}

Since the neutrinos are now known to have masses, the symmetry
considerations for the quark mass matrices are equally applicable
to the leptons. Furthermore, the leptons, not being confined,
have masses that are better defined than those of the quarks.
Replacing $d$ and $u$ by the charged leptons $\ell$ and the neutrinos
$\nu$, Eqs.~(\ref{Eq:md-nondiag}), (\ref{Eq:mu-nondiag}) and 
(\ref{Eq:b-vs-c}) take the form
\begin{eqnarray}
\label{Eq:mell-nondiag}
M(\ell)&=&\begin{pmatrix}
0    & d(\ell) & 0 \\
d(\ell) & c(\ell) & b(\ell) \\
0    & b(\ell) & a(\ell)
\end{pmatrix}, \\[3mm]
\label{Eq:nu-nondiag}
M(\nu)&=&\begin{pmatrix}
0    & id(\nu) & 0 \\
-id(\nu) & c(\nu) & b(\nu) \\
0    & b(\nu) & a(\nu)
\end{pmatrix},
\end{eqnarray}
with
\begin{equation}
\label{Eq:lept:b-vs-c}
b^2(\ell)=8c^2(\ell), \quad b^2(\nu)=8c^2(\nu).
\end{equation}

The rotation matrices $R(\ell)$ and $R(\nu)$ are defined in exactly
the same manner:
\begin{eqnarray}
\label{Eq:diagonal-ell}
M(\ell)&=&R(\ell)\text{diag}(m_e,-m_\mu,m_\tau)R^{\text{T}}(\ell), \\[3mm]
\label{Eq:diagonal-nu}
M(\nu)&=&\text{diag}(i,1,1)R(\nu)\text{diag}(m_1,-m_2,m_3)R^{\text{T}}(\nu)
\text{diag}(-i,1,1),
\end{eqnarray}
where in Eq.~(\ref{Eq:diagonal-nu}) the masses of the three neutrinos
are designated as $m_1$, $m_2$ and $m_3$.
The lepton CKM mixing matrix is
\begin{equation}
\label{Eq:lept:VKM-def}
V^\ell=R^{\text{T}}(\nu)\text{diag}(-i,1,1) R(\ell).
\end{equation}

There are actually significant differences between the quark case 
and the lepton case.
\begin{itemize}
\itemsep=-2pt
\item[(A)]
In the quark case, that there is $CP$ violation has been known for
many years \cite{Christenson:fg,Wu:qx}.
In the lepton case, it is not known whether $CP$ is conserved or not.
While it is tempting, on the basis of quark-lepton universality,
to believe that $CP$ non-conservation also holds for leptons,
the possibility of lepton $CP$ conservation cannot be excluded.
In the former case, Eq.~(\ref{Eq:nu-nondiag}) holds; 
in the latter case, both $i$ and $-i$ on the right-hand side
of Eq.~(\ref{Eq:nu-nondiag}) are replaced by 1.
To take both possibilities into account, Eq.~(\ref{Eq:nu-nondiag}) 
needs to be generalized to
\begin{equation}
\label{Eq:nu-nondiag-general}
M(\nu)=\begin{pmatrix}
0    & \epsilon d(\nu) & 0 \\
\epsilon^* d(\nu) & c(\nu) & b(\nu) \\
0    & b(\nu) & a(\nu)
\end{pmatrix},
\end{equation}
where the two cases above correspond to $\epsilon=i, 1$, respectively.
Similarly, Eqs.~(\ref{Eq:diagonal-nu}) and (\ref{Eq:lept:VKM-def}) take 
the forms
\begin{equation}
\label{Eq:diagonal-nu-general}
M(\nu)=\text{diag}(\epsilon,1,1)R(\nu)\text{diag}(m_1,-m_2,m_3)
R^{\text{T}}(\nu)\text{diag}(\epsilon^*,1,1),
\end{equation}
and
\begin{equation}
\label{Eq:lept:VKM-def-general}
V^\ell=R^{\text{T}}(\nu)\text{diag}(\epsilon^*,1,1) R(\ell).
\end{equation}

\item[(B)]
In the case of quarks, both the masses and the absolute values
of the elements of the CKM mixing matrix are known
experimentally. In the case of the leptons, there is much less 
experimental information.
While the masses of the three charged leptons are accurately known,
the corresponding knowledge about the neutrinos is limited to
differences between the masses squared, $m_1^2$, $m_2^2$, $m_3^2$.
It is for this reason that the lepton case may be considered to be
more challenging, and one of the first tasks is to determine
the individual masses from the existing experimental data.
This is to be carried out in Sec.~\ref{sec:9}.

\item[(C)]
The lack of knowledge about the neutrino masses has another profound
consequence. For quarks, the inequality (\ref{Eq:quark-mass-inequality})
holds for both $u$ and $d$.
In contrast, because of the presence of the minus sign with $m_2$
in Eq.~(\ref{Eq:diagonal-nu}), all that is known about neutrino
masses is
\begin{equation}
\label{Eq:nu-mass-inequality}
m_1\le m_3.
\end{equation}
In other words, for neutrinos, Eq.~(\ref{Eq:nu-mass-inequality}) can be
used, but not (\ref{Eq:quark-mass-inequality}).
The first task is therefore to determine the allowed region in the space
$(m_1,m_2,m_3)$ of neutrinos, which must be between those permitted
by (\ref{Eq:quark-mass-inequality}) and (\ref{Eq:nu-mass-inequality}).
\end{itemize}

It follows from (\ref{Eq:quarks:m1m2m3}) that the parameter $a$,
which must be positive, satisfies the cubic equation
\begin{equation}
\label{Eq:cubic-a}
9a^3-17S_1a^2+(8S_1^2+S_2)a-S_3=0.
\end{equation}
Any real cubic equation can have either one or three real solutions.
Where there is \textit{one} real solution, that one is negative,
and thus unphysical, as is seen from (\ref{Eq:quarks:m1m2m3}).
Where there are \textit{three} real solutions, one of them is negative,
while two are positive.
We shall refer to these two positive solutions as Solution 1 (larger
$a$) and Solution 2 (smaller $a$).

It is instructive to consider briefly the simple case $m_1=m_2=0$.
In this case, it follows from (\ref{Eq:quarks:m1m2m3}) that
$S_1=m_3$ and $S_2=S_3=0$, and (\ref{Eq:cubic-a}) reduces to
\begin{equation}
\label{Eq:cubic-a-simple}
9a^3-17m_3 a^2+8m_3^2 a=0,
\end{equation}
with the solutions
\begin{equation}
\label{Eq:limit:m1=m2=0}
a=m_3, \quad \frac{8}{9}\,m_3, \quad 0.
\end{equation}
Here 0 is the limiting value of the negative solution and hence is of
no interest. From the above definitions, in this case $a=m_3$ is
Solution~1 while $a=\frac{8}{9}\,m_3$ is Solution~2. From
(\ref{Eq:nu-nondiag-general}), the mass matrices are
\begin{equation}
\renewcommand{\arraycolsep}{.5em}
M(\nu)=\begin{pmatrix}
0  & 0 & 0 \\
0  & 0 & 0 \\
0  & 0 & 1
\end{pmatrix}m_3,
\end{equation}
for Solution~1, and
\begin{equation}
\renewcommand{\arraycolsep}{.5em}
M(\nu)=\begin{pmatrix}
0  & 0 & 0 \\
0  & \frac{1}{9} & \frac{2\sqrt{2}}{9} \\
0  & \frac{2\sqrt{2}}{9} & \frac{8}{9}
\end{pmatrix}m_3,
\end{equation}
for Solution~2.  

In Ref.~\cite{Lehmann:1995br} for the quark mass matrices, only
Solution~1 was considered.

The above considerations on the cubic equation (\ref{Eq:cubic-a}) can
be used to determine the allowed physical region in the $(m_1/m_3,
m_2/m_3)$ plane, as shown in Fig.~\ref{Fig:disc}.  This region is only
slightly larger than the triangle given by the inequality
(\ref{Eq:quark-mass-inequality}), with two additional regions, one
where $m_2>m_3$ and the other a very small one with
$m_1>m_2$.

\begin{figure}[t]
\refstepcounter{figure}
\label{Fig:disc}
\addtocounter{figure}{-1}
\begin{center}
\setlength{\unitlength}{1cm}
\begin{picture}(12,9.5)
\put(2.0,1.0)
{\mbox{\epsfysize=9.0cm\epsffile{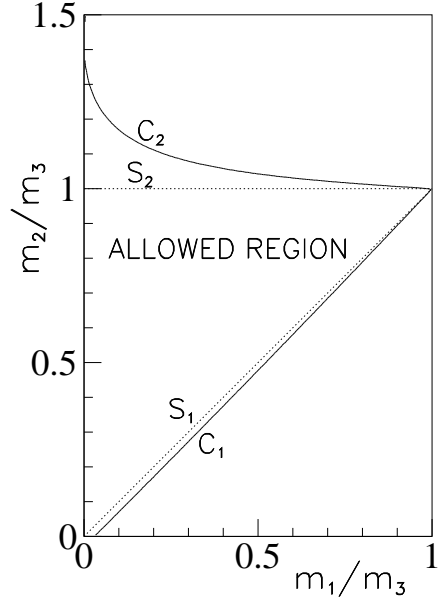}}}
\end{picture}
\vspace*{-8mm}
\caption{The allowed region for the three neutrino masses
$m_1$, $m_2$ and $m_3$ is bounded by parts of the axes and 
the \textit{solid} curves $\mathcal{C}_1$ and $\mathcal{C}_2$.
The lines $\mathcal{S}_1$ and $\mathcal{S}_2$ correspond to
$m_1=m_2$ and $m_2=m_3$, respectively.}
\end{center}
\end{figure}

The following notation is convenient.
\begin{itemize}
\itemsep=-2pt
\item[(1)]
$\mathcal{R}$ denotes the entire allowed region, not including the
boundary, for the three neutrino masses $m_1$, $m_2$ and $m_3$, as
shown in  Fig.~\ref{Fig:disc}.
\item[(2)]
$\mathcal{R}_0$ denotes the part  of $\mathcal{R}$  where $m_1<m_2<m_3$.
\item[(3)]
$\mathcal{R}_1$ denotes the part  of $\mathcal{R}$  where $m_1>m_2$.
\item[(4)]
$\mathcal{R}_2$ denotes the part  of $\mathcal{R}$  where $m_2>m_3$.
\item[(5)]
$\mathcal{S}_1$ denotes the part  of $\mathcal{R}$  where $m_1=m_2$;
similarly $\mathcal{S}_2$ the part where $m_2=m_3$.
\end{itemize}
Thus
\begin{equation}
\mathcal{R}= \mathcal{R}_0+ \mathcal{R}_1+ \mathcal{R}_2
+\mathcal{S}_1+\mathcal{S}_2.
\end{equation}
\begin{itemize}
\itemsep=-2pt
\item[(6)]
$\mathcal{C}_1$ is the curved part of the boundary of $\mathcal{R}_1$.
\item[(7)]
$\mathcal{C}_2$ is the curved part of the boundary of $\mathcal{R}_2$.
\item[(8)]
$\mathcal{C}_3$ is the plane
\begin{equation}
0 < m_2/m_3 < (17+\sqrt{33})/16\simeq 1.42, \quad m_1=0.
\end{equation}
\item[(9)]
$\mathcal{C}_4$ is the plane
\begin{equation}
0 < m_1/m_3 < 17-12\sqrt{2}\simeq 0.029, \quad m_2=0.
\end{equation}
\end{itemize}

\section{The Rotation Matrix}\label{sec:4}

There is an elementary but somewhat complicated issue of the sign 
ambiguities in the definition of the rotation matrices
$R(\nu)$ and $R(\ell)$ as given by 
Eqs.~(\ref{Eq:diagonal-ell}) and (\ref{Eq:diagonal-nu}).

It is seen from Eq.~(\ref{Eq:quarks:m1m2m3}) that, for either
Solution~1 or Solution~2 as defined in Sec.~\ref{sec:3}, the values of
$m_1$, $m_2$ and $m_3$ in $\mathcal{R}$ determine those of $a$, $b^2$ and
$d^2$, but not $b$ and $d$.  Therefore, for either $R(\nu)$ or
$R(\ell)$, there are actually eight distinct $R$'s in $\mathcal{R}$:
$R^{k\pm\pm}$, where $k=1$ for Solution~1 and $k=2$ for Solution~2, and
where the first $\pm$ and the second $\pm$ designate the ``parities''
(or sign factors) of $b$ and $d$ respectively (see below).

Since these $R$'s are determined by
\begin{equation}
\renewcommand{\arraycolsep}{.5em}
\begin{pmatrix}
0 & d & 0 \\
d & c & b \\
0 & b & a
\end{pmatrix}
R=R
\renewcommand{\arraycolsep}{.3em}
\begin{pmatrix}
m_1 & 0 & 0 \\
0 & -m_2 & 0 \\
0 & 0 & m_3
\end{pmatrix},
\end{equation}
the elements of $R$ are given explicitly by
\begin{alignat}{3}
\label{Eq:R-elements}
R_{11}&=d(m_1-a)y_1, &\quad
R_{12}&=-d(m_2+a)y_2, &\quad
R_{13}&=d(m_3-a)y_3, \nonumber \\
R_{21}&=m_1(m_1-a)y_1, &\quad
R_{22}&=m_2(m_2+a)y_2, &\quad
R_{23}&=m_3(m_3-a)y_3, \nonumber \\
R_{31}&=bm_1y_1, &\quad
R_{32}&=-bm_2y_2, &\quad
R_{33}&=bm_3y_3,
\end{alignat}
where the values of $y_1$, $y_2$ and $y_3$ are such that $R$ is
orthogonal.

This condition of orthogonality does not determine the signs of these
$y$'s:
\begin{equation}   \label{Eq:y_j}
y_j=\frac{\pm1}{\sqrt{d^2(\lambda_j-a)^2+\lambda_j^2(\lambda_j-a)^2
                   +b^2\lambda_j^2}},
\end{equation}
where
\begin{equation}
\lambda_1=m_1, \quad \lambda_2=-m_2, \quad \lambda_3=m_3.
\end{equation}
The entire problem is to choose the three $\pm$ signs in
Eq.~(\ref{Eq:y_j}).

Strictly speaking, any choice of sign will do.
Since such choices lead to a large number of possible
CKM mixing matrices for leptons
(for each $R$, there are $2^3=8$ possible choices of signs), 
it is useful to make an intelligent choice of these signs.

The basic principle to be used to choose the signs is that of 
\textit{continuity}, i.e., the continuity of the nine $R_{ij}$ for each $R$.
For example, each of these $R_{ij}$ must be continuous in $\mathcal{R}$.
Note that the continuity of an $R$ implies the continuity of its derivatives
with respect to the  masses $m_1$, $m_2$ and $m_3$.\looseness=-1

The problem to be solved is to find a set of eight $3\times 3$ matrices
$R^{k\pm\pm}$ with the following conditions:
\begin{itemize}
\itemsep=-2pt
\item[(1)]
$R^{k\pm\pm}$ are continuous in $\mathcal{R}$;
\item[(2)]
$R_{jj}^{k\pm\pm}>0$ in $\mathcal{R}_0$; and
\item[(3)]
$R^{1\pm\pm}=R^{2\pm\pm}$ on $\mathcal{C}_1$.
\end{itemize}

\begin{table}[h]
\begin{center}
\caption{Signs of the coefficients of the cubic equation
(\ref{Eq:cubic-c-simple}).}\label{tab:1}
\vspace{.175in}
\renewcommand{\tabcolsep}{.75em}
\begin{tabular}{|c|c|c|c|c|}
\hline 
&9&$-10S_1$&$S_1^2+S_2$&$-S_1S_2+S_3$\\
\hline
$\mathcal{R}_2$&$+$&$\pm$&$-$&$-$\\
$\mathcal{R}_0$&$+$&$-$&$\pm$&$+$\\
$\mathcal{R}_1$&$+$&$-$&$+$&$-$\\
\hline
 \end{tabular} 
\end{center}
\end{table}

\begin{table}[h]
\begin{center}
\caption{Signs of $c$ for Solution 1 and Solution
2.}\label{tab:2}
\vspace{.175in}
\renewcommand{\tabcolsep}{1em}
\begin{tabular}{|c|c|c|}
\hline 
&Solution 1&Solution 2\\
\hline
$\mathcal{R}_2$&$-$&$-$\\
$\mathcal{R}_0$&$-$&$+$\\
$\mathcal{R}_1$&$+$&$+$\\
\hline
 \end{tabular}
\vspace{-2pt}
\end{center}
\end{table}

Condition (3) requires the following comments. First, this condition can be
imposed on $\mathcal{C}_1$ \textit{or} $\mathcal{C}_2$, but not on
$\mathcal{C}_3$ or $\mathcal{C}_4$.  The reason is that, on $\mathcal{C}_1$
and $\mathcal{C}_2$, the values of $a$ for the first and second solutions are
the same.  Secondly, what this equation means is that, given an $R^1$, for
example $R^{1++}$, there is a choice for the two $\pm$ signs for $R^{2\pm\pm}$
so that the $R$'s are equal; this can only be achieved on \textit{either}
$\mathcal{C}_1$ \textit{or} $\mathcal{C}_2$, \textit{not on both}.  Actually,
there is no such choice: If $\mathcal{C}_2$ is chosen, no solution exists
because of the condition (2).  It is therefore necessary to impose this
condition on $\mathcal{C}_1$, for any ``$b$ parity'' and ``$d$ parity.'' This
is already indicated in condition (3) above.

The ``parities'' or sign factors are defined as follows: For positive
``$b$ parity,'' $b$ has the opposite sign of $c$, which is the convention
of \cite{Lehmann:1995br}. In particular, for Solution~1, $b$ is then
positive in $\mathcal{R}_0$. When the ``$b$ parity'' is flipped,
$R_{13}^{k}$, $R_{23}^{k}$, $R_{31}^{k}$ and $R_{32}^{k}$ change sign.
The ``$d$ parity'' is the sign of $d$ in all of $\mathcal{R}$. When this
sign is flipped, $R_{12}^{k}$, $R_{13}^{k}$, $R_{21}^{k}$ and
$R_{31}^{k}$ change sign.

Attention is next turned to the cubic equation (\ref{Eq:cubic-a-simple})
for $a$.  Since $c=S_1-a$ from Eq.~(\ref{Eq:quarks:m1m2m3}), the
corresponding cubic equation for $c$ is
\begin{equation}
\label{Eq:cubic-c-simple}
9c^3-10S_1c^2+(S_1^2+S_2)c-(S_1S_2-S_3)=0.
\end{equation}
The nice formula
\begin{equation}
S_3-S_1S_2=(m_2-m_1)(m_3-m_2)(m_3+m_1)
\end{equation}
shows that $S_3-S_1S_2$ is positive in $\mathcal{R}_0$, and in fact implies
that, in $\mathcal{R}_0$, two of the solutions $c$ from
Eq.~(\ref{Eq:cubic-c-simple}) are positive, while one is negative.  In a
similar way, the signs of the coefficients of the cubic equation
(\ref{Eq:cubic-c-simple}) are listed in Table~\ref{tab:1}.

Since the three solutions of this cubic equation are known to be all real
in $\mathcal{R}$, Table~\ref{tab:2} follows immediately from
Table~\ref{tab:1}. In particular
\begin{alignat}{2}
&c=0&\quad&\text{on } \mathcal{S}_1 \text{ for Solution 1}; \nonumber \\
&c=0&\quad&\text{on } \mathcal{S}_2 \text{ for Solution 2}.
\end{alignat}

The next task is to show that
\begin{equation} \label{Eq:a-m1}
a-m_1\ge0
\end{equation}
in $\mathcal{R}$, where the equality sign holds only on $\mathcal{S}_2$,
and then only for Solution~2. Similarly,
\begin{equation} \label{Eq:m3-a}
m_3-a\ge0
\end{equation}
in $\mathcal{R}$, where the equality sign holds only on $\mathcal{S}_1$,
and then only for Solution~1.
These derivations are straightforward and hence omitted. Note that the
left-hand sides of these inequalities play an important role  in
Eq.~(\ref{Eq:R-elements}). The two relations may conveniently be
summarized as\vadjust{\kern-8pt}
\begin{equation}
m_1\underset{\mathcal{S}_2, \text{Sol }2}{\le} a 
   \underset{\mathcal{S}_1, \text{Sol }1}{\le} m_3.
\end{equation}
It is also useful to note that
\begin{equation}
m_1 \underset{\mathcal{S}_1, \text{Sol }1}{=} |d|, \quad
|d| \underset{\mathcal{S}_2, \text{Sol }2}{=} m_3,
\end{equation}
but in this latter case, these are not extrema.

\begin{figure}[t]
\refstepcounter{figure}
\label{Fig:Rnu-cont-1}
\addtocounter{figure}{-1}
\begin{center}
\setlength{\unitlength}{1cm}
\begin{picture}(12,12)
\put(-1.0,-3.0)
{\mbox{\epsfysize=15.0cm\epsffile{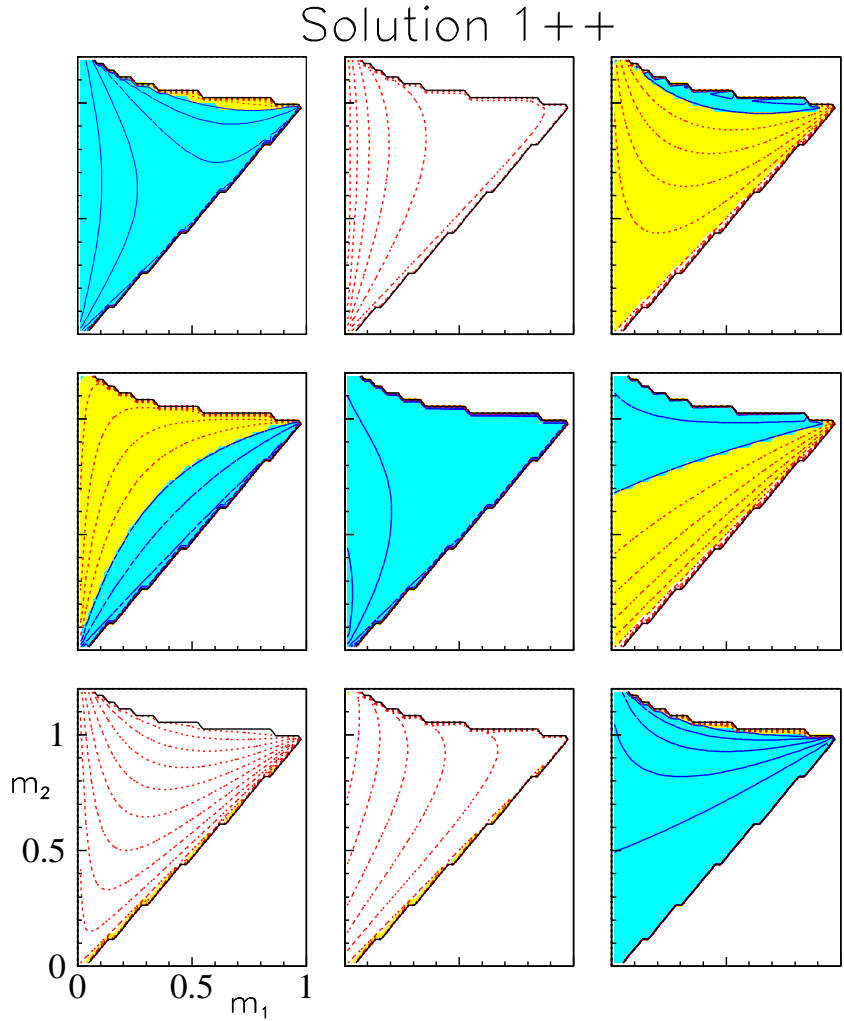}}}
\end{picture}
\vspace*{25mm}
\caption{Solution 1: $R^{1++}_{ij}(\nu)$ matrix elements \textit{vs.}\ $m_1$ 
and $m_2$ ($m_3=1$), for $i=1,2,3$ and $j=1,2,3$.
The dark regions indicate where the values exceed 0.5, and the lightly
shaded regions indicate where the values are between 0 and 0.5.  In the
unshaded (white) regions, the values are negative.}
\end{center}
\end{figure}

\begin{figure}[t]
\refstepcounter{figure}
\label{Fig:Rnu-cont-2}
\addtocounter{figure}{-1}
\begin{center}
\setlength{\unitlength}{1cm}
\begin{picture}(12,12)
\put(-1.0,-3.0)
{\mbox{\epsfysize=15.0cm\epsffile{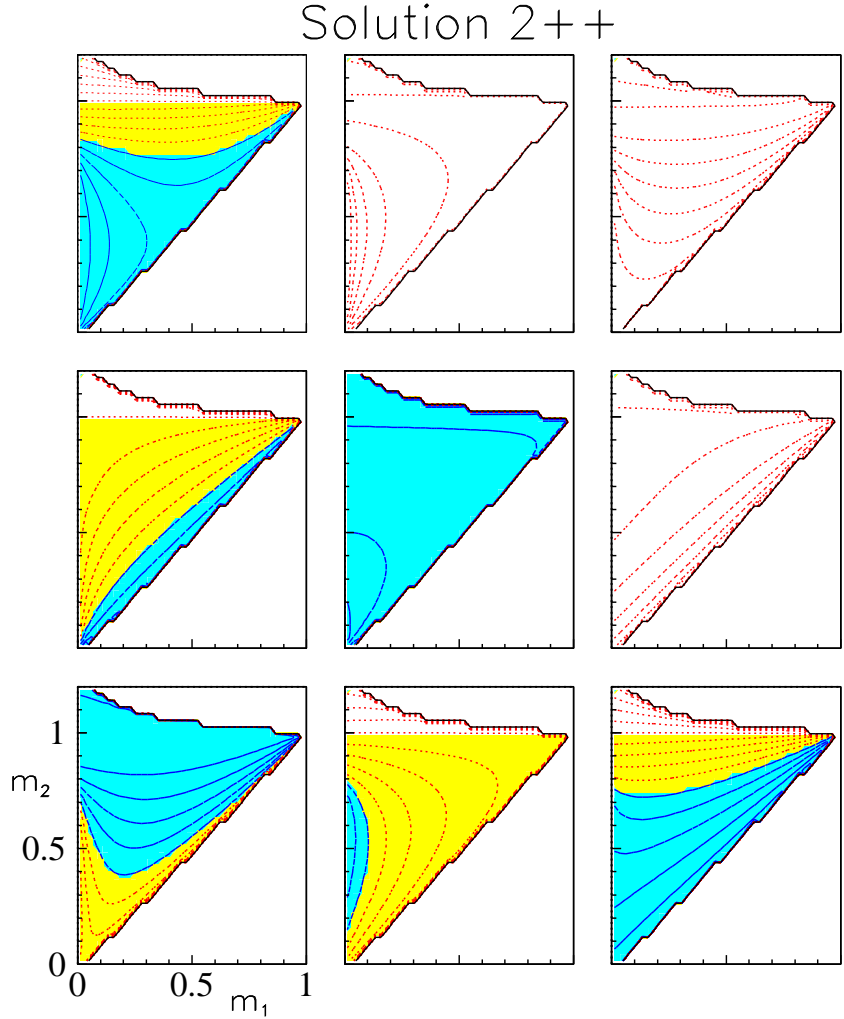}}}
\end{picture}
\vspace*{25mm}
\caption{Solution 2: $R^{2++}_{ij}(\nu)$ matrix elements \textit{vs.}\ $m_1$ 
and $m_2$ ($m_3=1$).
Compare with Fig.~\ref{Fig:Rnu-cont-1}.}
\end{center}
\end{figure}

With this knowledge, the resulting signs are easily determined as given in
Figs.~\ref{Fig:Rnu-cont-1} and \ref{Fig:Rnu-cont-2}. The signs of $y_j$,
respecting conditions (1), (2) and (3) above, is such that:
\begin{itemize}
\itemsep=0pt
\item
For Solution~1, some $R_{ij}^{1\pm\pm}$ vanish on $\mathcal{S}_1$. They are
$ij=13, 23, 31$ and 32, corresponding to $m_3-a$ and $b$ becoming zero.
(These $R_{ij}^{1\pm\pm}$ thus have opposite signs in $\mathcal{R}_0$ and
$\mathcal{R}_1$.)
\item
For Solution~2, some $R_{ij}^{2\pm\pm}$ vanish on $\mathcal{S}_2$.
They are $ij=11, 21, 32$ and 33, corresponding to $a-m_1$ and $b$
becoming zero.
(These $R_{ij}^{2\pm\pm}$ thus have opposite signs in $\mathcal{R}_0$
and $\mathcal{R}_2$.)
\end{itemize}
In order to have $R_{ij}^{k\pm\pm}$ that are continuous, it is required to
flip signs of some $y_j$ as these boundaries $\mathcal{S}_1$ and
$\mathcal{S}_2$ are crossed.

The mixing matrices display strong variations with $m_1$ and $m_2$. For
Solution~1, the diagonal elements dominate in much of the parameter space,
whereas for Solution~2, this is not the case.

For the charged leptons, the masses are strongly hierarchical. Thus, the
rotation matrices correspond to the lower left-hand corners of those
displayed in Figs.~\ref{Fig:Rnu-cont-1} and \ref{Fig:Rnu-cont-2}.  For
Solution~1, this is close to the unit matrix, whereas for Solution~2
certain non-diagonal elements are also significant.

\section{The Mixing Matrix}\label{sec:5}

We write the unitary (but not necessarily real) mixing matrix as
\begin{equation}
\renewcommand{\arraycolsep}{.5em}
U=\left[
\begin{array}{ccc}
U_{e1} & U_{e2} & U_{e3} \\
U_{\mu 1} & U_{\mu 2} & U_{\mu 3} \\
U_{\tau 1} & U_{\tau 2} & U_{\tau 3}
\end{array}\right],
\end{equation}
where (cf. Eq.~(\ref{Eq:lept:VKM-def-general}))
\begin{equation}
U=(V^{\ell})^\dagger
=R^{\text{T}}(\ell)\,\text{diag}(\epsilon,1,1)R(\nu)
\end{equation}
relates the neutrino mass eigenstates to the flavor states:
\begin{equation}
|\nu_e\rangle=U_{e1}|\nu_1\rangle+U_{e2}|\nu_2\rangle+U_{e3}|\nu_3\rangle,
\end{equation}
etc. These are the states which enter in charged-current interactions.

\begin{figure}[htb]
\refstepcounter{figure}
\label{Fig:U21mp-cont-1}
\addtocounter{figure}{-1}
\begin{center}
\setlength{\unitlength}{1cm}
\begin{picture}(12,12)
\put(-1.0,-3.0)
{\mbox{\epsfysize=15.0cm\epsffile{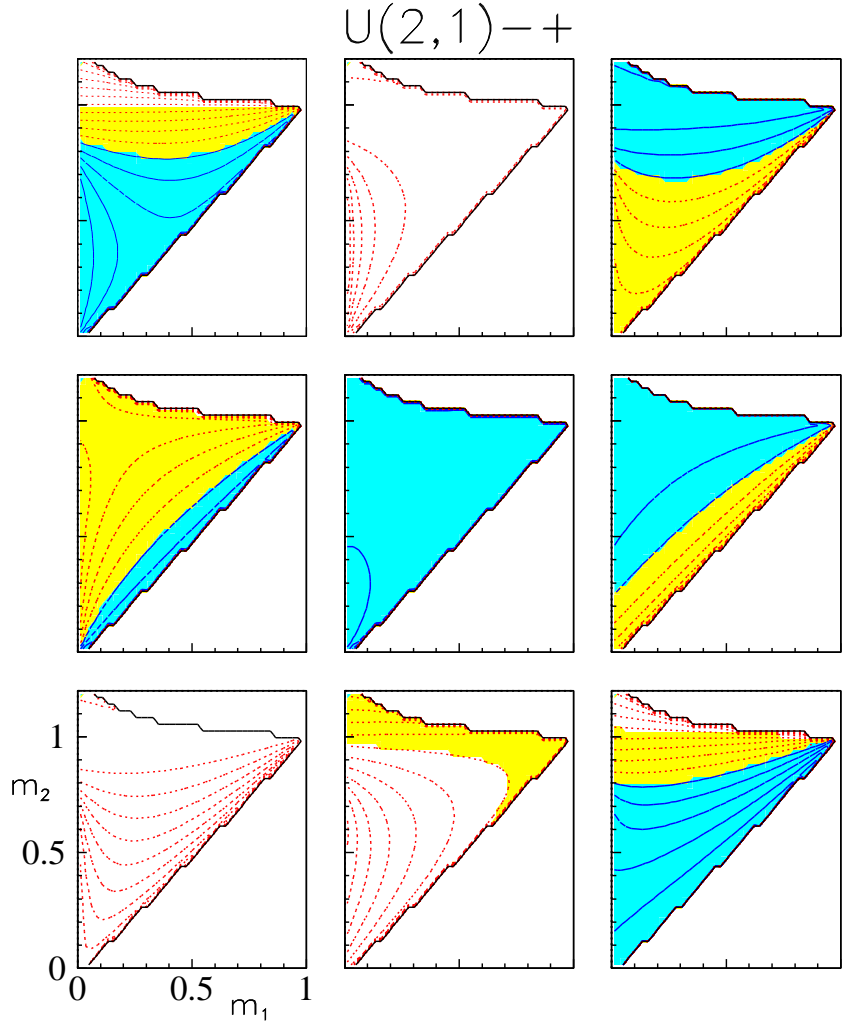}}}
\end{picture}
\vspace*{25mm}
\caption{Mixing matrix $[U(2,1)-+]_{ij}$ \textit{vs.}\ $m_1$ 
and $m_2$ ($m_3=1$), for $i=1,2,3$ and $j=1,2,3$.
The dark regions indicate where the values exceed 0.5, and the lightly
shaded regions indicate where the values are between 0 and 0.5.  In the
unshaded (white) regions, the values are negative.}
\end{center}
\end{figure}

We shall label these mixing matrices by the Solutions (1 or 2) involved 
in the rotation matrices $R(\nu)$ and $R(\ell)$, and the ``parities''
associated with the parameters $b$ and $d$, as discussed in
Sec.~\ref{sec:4}. Actually, only the product of the ``$b$-parities'' and
that of the ``$d$-parities'' matter. Thus it is convenient to define
\begin{align}
b\text{ parity}&=(b\text{ parity})_\nu\times(b\text{ parity})_\ell,
\nonumber \\
d\text{ parity}&=(d\text{ parity})_\nu\times(d\text{ parity})_\ell.
\end{align}
Therefore, there are a total of
\begin{equation}
2\times2\times2\times2\times2
\end{equation}
different $U$ matrices, where two factors of 2 arise from the two
Solutions for the $\nu$ and $\ell$ sectors, two factors of 2 arise from
the ``$b$ parities'' and the ``$d$ parities,'' and the last factor of 2
arises from the choice of $CP$ conservation or $CP$ non-conservation
($\epsilon=1$ or $\epsilon=i$).

For a representative case, Solution~2 for $R(\nu)$ and Solution~1 for
$R(\ell)$, $b\text{ parity}=-$, $d\text{ parity}=+$, and no $CP$
violation, we show in Fig.~\ref{Fig:U21mp-cont-1} the resulting mixing 
matrix $U(2,1)-+$. Since, for the charged leptons, Solution~1 is close to
the unit matrix, the resulting $U(2,1)-+$ is rather similar to the
corresponding $R(\nu)$ for Solution~2 and parities $-+$.

\section{Atmospheric Neutrinos}\label{sec:6}

Let us next review the oscillation formulas that are relevant for
atmospheric neutrinos. A neutrino state, which was a pure $\nu_\alpha$
state at $t=0$, will at time $t$ have the form
\begin{equation}
\nu_\alpha(t) = \sum_{j=1}^3 U_{\alpha j}e^{-iE_jt}\nu_j \qquad
(\alpha=e,\mu,\tau)\; .
\end{equation}
For relativistic neutrinos, the energy is given as $E_j\simeq
p+m_j^2/2E$, and the survival probability for the $\nu_\mu$ takes the form
\begin{eqnarray}
\label{Eq:atm-survival}
P_{\nu_\mu\rightarrow \nu_\mu}(t)
&=& 1
-4\biggl[|U_{\mu1}|^2|U_{\mu2}|^2
\sin^2\left(\frac{\Delta m^2_{21}t}{4E}\right)
+|U_{\mu1}|^2|U_{\mu3}|^2
\sin^2\left(\frac{\Delta m^2_{31}t}{4E}\right)
\nonumber \\
&&
\phantom{1-4\biggl[}
+|U_{\mu2}|^2|U_{\mu3}|^2
\sin^2\left(\frac{\Delta m^2_{32}t}{4E}\right)
\biggr],
\end{eqnarray}
where $\Delta m_{ij}^2 = m_{i}^2 - m_{j}^2$.

After propagation over some distance, a neutrino of a different flavor may
appear. The probability amplitude for the transition $\alpha\to\beta$ is
given by
\begin{equation} \label{Eq:trans-prob}
\langle\nu_\beta(t)|\nu_\alpha(0)\rangle
=\sum_k U_{\beta k}^* U_{\alpha k}\, e^{i m_k^2 t/2E}.
\end{equation}
In contrast to the survival probability, this expression is not invariant
under complex conjugation of $U$.

A case of particular interest is
\begin{eqnarray}
\label{Eq:nu-to-tau}
P_{\nu_\mu\rightarrow \nu_\tau}(t)
&=&
\big|\sum_k U_{\tau k}^* U_{\mu k}\, e^{i m_k^2 t/2E}\big|^2 \nonumber \\
&=&\bigg\{-4\,\text{Re}[U_{\tau 1}^* U_{\mu 2}^* U_{\tau 2} U_{\mu 1}]\,
\sin^2\bigg(\frac{\Delta m_{21}^2 t}{4E}\bigg) \nonumber \\
& & \quad
\mbox{} -2\,\text{Im}[U_{\tau 1}^* U_{\mu 2}^* U_{\tau 2} U_{\mu 1}]\,
\sin\bigg(\frac{\Delta m_{21}^2 t}{2E}\bigg)\bigg\} + \text{cyclic},
\end{eqnarray}
where we have used the orthogonality, $\sum_k U_{\tau k}^* U_{\mu k}=0$.
For the real case, this simplifies to
\begin{eqnarray}
\label{Eq:nu-to-tau-real}
P_{\nu_\mu\rightarrow \nu_\tau}(t)
&=&
-4U_{\tau1}U_{\mu2}U_{\tau2}U_{\mu1}
\sin^2\left(\frac{\Delta m^2_{21}t}{4E}\right)
-4U_{\tau2}U_{\mu3}U_{\tau3}U_{\mu2}
\sin^2\left(\frac{\Delta m^2_{32}t}{4E}\right)
\nonumber \\
&&\mbox{} -4U_{\tau3}U_{\mu1}U_{\tau1}U_{\mu3}
\sin^2\left(\frac{\Delta m^2_{13}t}{4E}\right)\, .
\end{eqnarray}

It is instructive to study the simple limit
\begin{equation}
\label{Eq:small-m21}
|\Delta m^2_{21}| \ll |\Delta m^2_{32}|, \quad
\Delta m^2_{21}t/4E \ll 1,
\end{equation}
with all $U_{\alpha k}=\mathcal{O}(1)$.
Then, by unitarity, Eq.~(\ref{Eq:nu-to-tau-real}) simplifies to the
familiar expression
\begin{equation}
\label{Eq:nu-to-tau-simple}
P_{\nu_\mu\rightarrow \nu_\tau}(t)
\simeq 4|U_{\mu3}|^2|U_{\tau3}|^2\,
\sin^2\left(\frac{\Delta m^2_{32}t}{4E}\right).
\end{equation}

Fitting the data within a two-flavor model, with $\sin^2\theta$ and
$\Delta m^2$ as independent parameters, one finds \cite{Fukuda:1998tw}
large mixing angles. In the limit of Eq.~(\ref{Eq:small-m21}), this
corresponds to large values for $|U_{\mu3}U_{\tau3}|^2$, see
Eq.~(\ref{Eq:nu-to-tau-simple}).

The observed suppression of atmospheric $\nu_\mu$ \cite{Fukuda:1998tw}
suggests masses of the order $m\simeq0.05$~eV. In order to determine the
neutrino masses, we formulate a $\chi^2$ by comparing predicted $\nu_\mu$
and $\nu_e$ fluxes with data:
\begin{equation}
\chi^2_{\text{atm}}=\sum_i\frac{(\Phi_i-\Phi_i^{\text{exp}})^2}{\sigma_i^2}.
\end{equation}
The experimental data used are those from Super-Kamiokande
\cite{Fukuda:1998tw}: the 8 data points (bins in $E/L$, where $L=ct$) for
$\nu_\mu$ and the 8 data points for $\nu_e$. These sixteen data points are
treated as separate inputs,  allowing an overall normalization constant
for the two sets of data. Also, since the various survival and transition
probabilities are rather sensitive to the precise values of energy and
oscillation length, we averaged over these, within each of the 8 bins.

We show in Fig.~\ref{Fig:lvls-05} the contributions to $\chi^2$ from the
atmospheric-neutrino data, for the mixing matrices $U(2,1)-+$
corresponding to $CP$ conservation (left part) and $CP$ non-conservation
(right part). The figure shows $\chi^2$ as a function of $m_1$ and $m_2$,
for fixed $m_3=0.05\text{ eV}$.

The different solutions and parities that determine the mixing matrices give
rather differently shaped $\chi^2$ minima when plotted \textit{vs.}\ $m_1$ and
$m_2$. For most cases, the minima occur inside the region $\mathcal{R}_0$. For
others, they occur near $\mathcal{S}_1$ or near $\mathcal{S}_2$.

Comparing with Figs.~\ref{Fig:Rnu-cont-1} and \ref{Fig:Rnu-cont-2}, we see
that Solution~1 provides large mixing for $m_2$ being a sizable fraction
of $m_3$, whereas Solution~2 favors relatively smaller values of $m_2$, or
$m_2$ close to $m_1$.

\section{The Three-Family MSW Mechanism}\label{sec:7}

The coupled equations satisfied by the three neutrino wave functions are
\cite{Wolfenstein:1977ue}
\begin{equation}
\label{Eq:Schr-1}
\renewcommand{\arraycolsep}{.375em}
i\,\frac{d}{d r}
\begin{pmatrix}
\phi_1(r) \\ \phi_2(r) \\ \phi_3(r)
\end{pmatrix}
=\left[
\begin{pmatrix}
D(r) & 0 & 0 \\
0 & 0 & 0 \\
0 & 0 & 0 \\
\end{pmatrix}
+\frac{1}{2p}
\begin{pmatrix}
M^2_{11} & M^2_{12} & M^2_{13}\\
M^2_{21} & M^2_{22} & M^2_{23}\\
M^2_{31} & M^2_{32} & M^2_{33}
\end{pmatrix}
\right]
\begin{pmatrix}
\phi_1(r) \\ \phi_2(r) \\ \phi_3(r)
\end{pmatrix},
\end{equation}
where
$D(r)=\sqrt{2}\,G_{\text{F}} N_e(r)$,
with $G_{\text{F}}$ the Fermi weak-interaction constant and $N_e(r)$ the
solar electron density at a distance $r$ from the center of the sun.
Furthermore, we denote the flavor states $\nu_e=\phi_1$,
$\nu_\mu=\phi_2$, $\nu_\tau=\phi_3$. These are the states which enter in
charged-current interactions.

The evolution of the neutrino wave functions is determined by the squared
mass matrix,
\begin{equation}   \label{Eq:M-squared}
\renewcommand{\arraycolsep}{.5em}
[M(\nu)]^2
=\begin{pmatrix}
d^2 & \epsilon cd & \epsilon bd \\
\epsilon^* cd & b^2+c^2+d^2 & b(a+c) \\
\epsilon^* bd & b(a+c) & a^2+b^2
\end{pmatrix}
\equiv\begin{pmatrix}
M^2_{11} & M^2_{12} & M^2_{13}\\
M^2_{21} & M^2_{22} & M^2_{23}\\
M^2_{31} & M^2_{32} & M^2_{33}
\end{pmatrix},
\end{equation}
the neutrino momentum, $p$,
and the solar electron density.
Here, $M^2_{ij}\equiv(M^2)_{ij}$, and $\epsilon=1$ ($CP$ conservation)
or $\epsilon=i$ ($CP$ non-conservation).

It is actually a good approximation to take an exponential electron
density, $N_e(r)=N_e(0)\exp(-r/r_0)$.
A fit to the solar density as given by \cite{Bahcall:2000nu} gives
$r_0=6.983\times10^4$~km.
For this case of an exponential solar density, the three-component
wave equation can be solved in terms of generalized
hypergeometric functions, ${}_2F_2$ \cite{Osland:1999et}.

The case treated in \cite{Osland:1999et} was that of a real mass matrix.
In that case, by scaling and shifting the radial variable, $u=r/r_0+u_0$,
with $u_0$ determined such that
\begin{equation}
D(0)r_0e^{u_0}=1,
\end{equation}
Eq.~(\ref{Eq:Schr-1}) could be transformed into the form
\begin{equation}
\label{Eq:Schr-4}
\renewcommand{\arraycolsep}{.5em}
i\,\frac{d}{du}
\begin{bmatrix}
\psi_1(u)\\ \psi_2(u)\\ \psi_3(u)
\end{bmatrix}
=
\begin{bmatrix}
\omega_1+e^{-u} & \chi_2 & \chi_3\\
\chi_2 & \omega_2 & 0\\
\chi_3 & 0 & \omega_3
\end{bmatrix}
\begin{bmatrix}
\psi_1(u)\\ \psi_2(u)\\ \psi_3(u)
\end{bmatrix},
\end{equation}
with $\omega_1$, $\omega_2$, $\omega_3$, $\chi_2$ and $\chi_3$ all real.

We now have to address a small complication due to the possible 
non-reality of the mass matrix induced by $CP$ non-conservation,
and the fact that also the charged lepton states have to be rotated.
Consider the case of $CP$ non-conservation, i.e., $\epsilon=i$.
In order to follow as closely as possible the procedure of 
\cite{Osland:1999et}, we need to rotate to a neutrino basis which in 
the absence of matter ($D(r)=0$) becomes that of the mass eigenstates.
This now involves diagonalizing the lower right-hand part not of 
$[M(\nu)]^2$, Eq.~(\ref{Eq:M-squared}), but of
\begin{equation}
\renewcommand{\arraycolsep}{.5em}
N^2=R^{\text{T}}(\ell) M^2 R(\ell)
=\begin{bmatrix}
(N^2)_{11} & (N^2)_{12} & (N^2)_{13}\\
(N^2)_{12}^* & (N^2)_{22} & (N^2)_{23}\\
(N^2)_{13}^* & (N^2)_{23}^* & (N^2)_{33}
\end{bmatrix},
\end{equation}
where not only the (1,2) and (1,3) elements are complex 
[together with (2,1) and (3,1)] (cf.\ Eq.~(\ref{Eq:M-squared})), 
but also the (2,3) and (3,2) elements.
This prevents a simple diagonalization like in \cite{Osland:1999et}.

However, we may make the lower right-hand part real by a unitary $2\times2$
rotation $\mathcal{U}$. In other words, we perform the transformation
\begin{equation}
\renewcommand{\arraycolsep}{.5em}
\mathcal{U}
\begin{bmatrix}
(N^2)_{22} & (N^2)_{23}\\
(N^2)_{23}^* & (N^2)_{33}
\end{bmatrix}
\mathcal{U}^\dagger
=\begin{bmatrix}
x & y \\
y & z
\end{bmatrix}, \quad \text{with }
\begin{bmatrix}
\bar\phi_2 \\ \bar\phi_3
\end{bmatrix}
=\mathcal{U}
\begin{bmatrix}
\phi_2 \\ \phi_3
\end{bmatrix}.
\end{equation}
By also defining  
\begin{equation}
\renewcommand{\arraycolsep}{.5em}
\begin{bmatrix}
\cos\theta_0 & -\sin\theta_0\\
\sin\theta_0 & \cos\theta_0
\end{bmatrix}
\frac{r_0}{2p}
\begin{bmatrix}
x & y \\
y & z
\end{bmatrix}
\begin{bmatrix}
\cos\theta_0 & \sin\theta_0\\
-\sin\theta_0 & \cos\theta_0
\end{bmatrix}
=
\begin{bmatrix}
\omega_2 & 0\\
0 & \omega_3
\end{bmatrix}
\end{equation}
together with
\begin{equation}
\renewcommand{\arraycolsep}{.5em}
\omega_1=\frac{r_0}{2p}\,(N^2)_{11}, \quad
\begin{bmatrix}
\chi_2\\ \chi_3
\end{bmatrix}
=\frac{r_0}{2p}
\begin{bmatrix}
\cos\theta_0 & -\sin\theta_0\\
\sin\theta_0 & \cos\theta_0
\end{bmatrix}
\begin{bmatrix}
(N^2)_{12} \\ (N^2)_{13}
\end{bmatrix}
\end{equation}
and
\begin{equation}
\renewcommand{\arraycolsep}{.5em}
\psi_1(u)=\phi_1(u), \quad
\begin{bmatrix}
\psi_2(u) \\ \psi_3(u)
\end{bmatrix}
=
\begin{bmatrix}
\cos\theta_0 & -\sin\theta_0\\
\sin\theta_0 & \cos\theta_0
\end{bmatrix}
\begin{bmatrix}
\bar\phi_2(u)\\
\bar\phi_3(u)
\end{bmatrix},
\end{equation}
then Eq.~(\ref{Eq:Schr-1}) can be written in the desired form
(\ref{Eq:Schr-4}). The $\omega$'s are real, but the $\chi_2$ and $\chi_3$
will in general be complex.
The rotations among $\phi_2$ and $\phi_3$ given by $\mathcal{U}$ and
$\theta_0$ need not concern us, since we are here only interested
in the electron neutrino, $\nu_e$.

When $\omega_2\ne\omega_3$, these $\psi_1$, $\psi_2$ and $\psi_3$
can be expressed uniquely in terms of a single $\psi$:
\begin{eqnarray}
\label{Eq:psi1-2-3}
\psi_1&=&\left(i\,\frac{d}{du}-\omega_2\right)
\left(i\,\frac{d}{du}-\omega_3\right)\psi, \nonumber \\
\psi_2&=&\chi_2\left(i\,\frac{d}{du}-\omega_3\right)\psi, \nonumber \\
\psi_3&=&\chi_3\left(i\,\frac{d}{du}-\omega_2\right)\psi,
\end{eqnarray}
where $\psi$ satisfies the third-order ordinary differential
equation \cite{Osland:1999et}
\begin{eqnarray}
\label{Eq:Schr-5}
\lefteqn{\left(i\,\frac{d}{du}-\mu_1\right)
\left(i\,\frac{d}{du}-\mu_2\right)
\left(i\,\frac{d}{du}-\mu_3\right)\psi}\qquad\nonumber \\
&&\mbox{}=e^{-u}
\left(i\,\frac{d}{du}-\omega_2\right)
\left(i\,\frac{d}{du}-\omega_3\right)\psi.
\end{eqnarray}
Here, $\mu_1$, $\mu_2$, and $\mu_3$ are the eigenvalues of
the right-hand matrix in Eq.~(\ref{Eq:Schr-4}) (without
the term $e^{-u}$), ordered such that
\begin{equation}
\mu_1 \le \mu_2 \le \mu_3.
\end{equation}

Equation (\ref{Eq:Schr-5}) is the differential equation 
for the generalized hypergeometric function $_2F_2$---see, 
for example, p.~184 of \cite{Bateman}.
Three linearly independent solutions of this third-order differential
equation (\ref{Eq:Schr-5}) are\hfil\eject
\vspace*{-2.5\baselineskip}
\begin{eqnarray}
\psi^{(1)}
&=&K_1e^{-i\mu_1u}
{}_2F_2\left[\left.
\begin{matrix}
-i(\omega_2-\mu_1), & -i(\omega_3-\mu_1)\\[4pt]
1-i(\mu_2-\mu_1),    & 1-i(\mu_3-\mu_1)
\end{matrix}
\right|ie^{-u} \right], \nonumber \\[6pt]
\psi^{(2)}
&=&K_2e^{-i\mu_2u}
{}_2F_2\left[\left.
\begin{matrix}
-i(\omega_2-\mu_2), & -i(\omega_3-\mu_2)\\[4pt]
1-i(\mu_1-\mu_2),    & 1-i(\mu_3-\mu_2)
\end{matrix}
\right|ie^{-u} \right], \nonumber \\[6pt]
\psi^{(3)}
&=&K_3e^{-i\mu_3u}
{}_2F_2\left[\left.
\begin{matrix}
-i(\omega_2-\mu_3), & -i(\omega_3-\mu_3)\\[4pt]
1-i(\mu_1-\mu_3),    & 1-i(\mu_2-\mu_3)
\end{matrix}
\right|ie^{-u} \right],
\end{eqnarray}
where \textit{K}, $K_1$, $K_2$ and $K_3$ are arbitrary non-zero constants.
Since Eq.~(\ref{Eq:Schr-5}) is linear, the general solution is
\begin{equation}
\psi=C_1\psi^{(1)}+C_2\psi^{(2)}+C_3\psi^{(3)},
\end{equation}
from which the $\psi_1$, $\psi_2$ and $\psi_3$ can be obtained
using Eq.~(\ref{Eq:psi1-2-3}).

For the case of two flavors, the products in (\ref{Eq:Schr-5}) consist
of one less term each,
and a familiar confluent hypergeometric function ${}_1F_1$
(also denoted Whittaker function or parabolic cylinder function)
is obtained \cite{Petcov:1987zj}.

These functions are trivial when $u\to\infty$. In fact, outside the sun,
they can be approximated by the exponential prefactors, since
\begin{equation}
\renewcommand{\arraycolsep}{.5em}
{}_2F_2\biggl[
\begin{matrix}
a_1, & a_2\\
b_1, & b_2
\end{matrix}\ 
\bigg|\ 0\biggr]=1.
\end{equation}

In order to impose the boundary conditions that only \textit{electron}
neutrinos are produced in the sun, we have to determine these functions
at large and negative values of $u$.
The series expansion is in principle convergent, but it is not practical
for large absolute values of both parameters and the argument.
One possible way of dealing with these generalized hypergeometric
functions has been given in \cite{Osland:1999et}.
The procedure used there is as follows.
First, $\psi_1^{(3)}$, $\psi_2^{(3)}$ and $\psi_3^{(3)}$ are evaluated
approximately using Barnes' integral representation for
${}_2F_2$ \cite{Bateman}.
Since we have not managed to apply this same procedure to the other
$\psi$'s, they are expressed in terms of another generalized
hypergeometric function ${}_3F_1$.
Since ${}_3F_1$ has an integral representation in terms of the
usual hypergeometric function ${}_2F_1$, these ${}_3F_1$ can be
evaluated by numerical integration.
The choice of the contours of integration has been discussed
in detail in \cite{Osland:1999et}.

For completeness we give in Appendix~\ref{app:B} asymptotic formulas for these
${}_3F_1$.  These asymptotic formulas turn out to be quite useful and in
particular are accurate for the region of the minimum $\chi^2$, to be
discussed below.

Some details on the book-keeping of reconstructing the neutrino
wave functions from the ${}_2F_2$ and ${}_3F_1$ are given in
Appendix~\ref{app:C}.

\section{Solar Neutrinos}\label{sec:8}

In order to compare the predictions of the model to data, we form a
$\chi^2$ by comparing the predictions to the available flux data.  For the
solar-neutrino flux, we take the values given by the `BP00' solar model
\cite{Bahcall:2000nu}.  For the solar-neutrino data, we use the total rates
from the Chlorine experiment \cite{Cleveland:1994er}, the Gallium
experiments \cite{Abdurashitov:1999bv,Hampel:1998xg} (we average the two
results), the Super-Kamiokande experiment \cite{Fukuda:1998fd}, and the
SNO experiment \cite{Ahmad:2001an,Ahmad:2002jz}.  We adopt the neutrino
energy spectra and detector efficiencies as given by Bahcall \textit{et al.}
\cite{Bahcall:1996qv}, and, for Super-Kamiokande and SNO, we also include
the neutral-current cross section \cite{Fukuda:1998tw}.  We do not
consider the day-night effect, since this is consistent with zero
\cite{Fukuda:1998fd}. Neither do we consider the electron-recoil spectrum,
since this is consistent with being flat \cite{Fukuda:1998ua} (see,
however \cite{Osland:2000gi}).  For the solar flux, we integrate over the
spectrum $\Phi_j(E_\nu)$, taking into account the detector efficiency
$\epsilon(E_\nu)$:
\begin{equation}
\Phi=\sum_j\int dE_\nu\, \Phi_j(E_\nu)\epsilon(E_\nu)P_\nu(E_\nu).
\end{equation}

In Fig.~\ref{Fig:lvls-05} we show the contributions to $\chi^2$ from the
solar-neutrino data, as functions of $m_1$ and $m_2$, for $m_3 = 0.05$~eV. 
As opposed to the atmospheric-neutrino data, the solar-neutrino data give
a minimum $\chi^2$ that is well localized in the $m_1$--$m_2$ plane, with
little dependence on $m_3$.

\begin{figure}
\refstepcounter{figure}
\label{Fig:lvls-05}
\addtocounter{figure}{-1}
\begin{center}
\vspace{.7in}
\setlength{\unitlength}{1cm}
\begin{picture}(15.5,18.8)
\put(1.,13.0){
\mbox{\epsfysize=7.5cm\epsffile{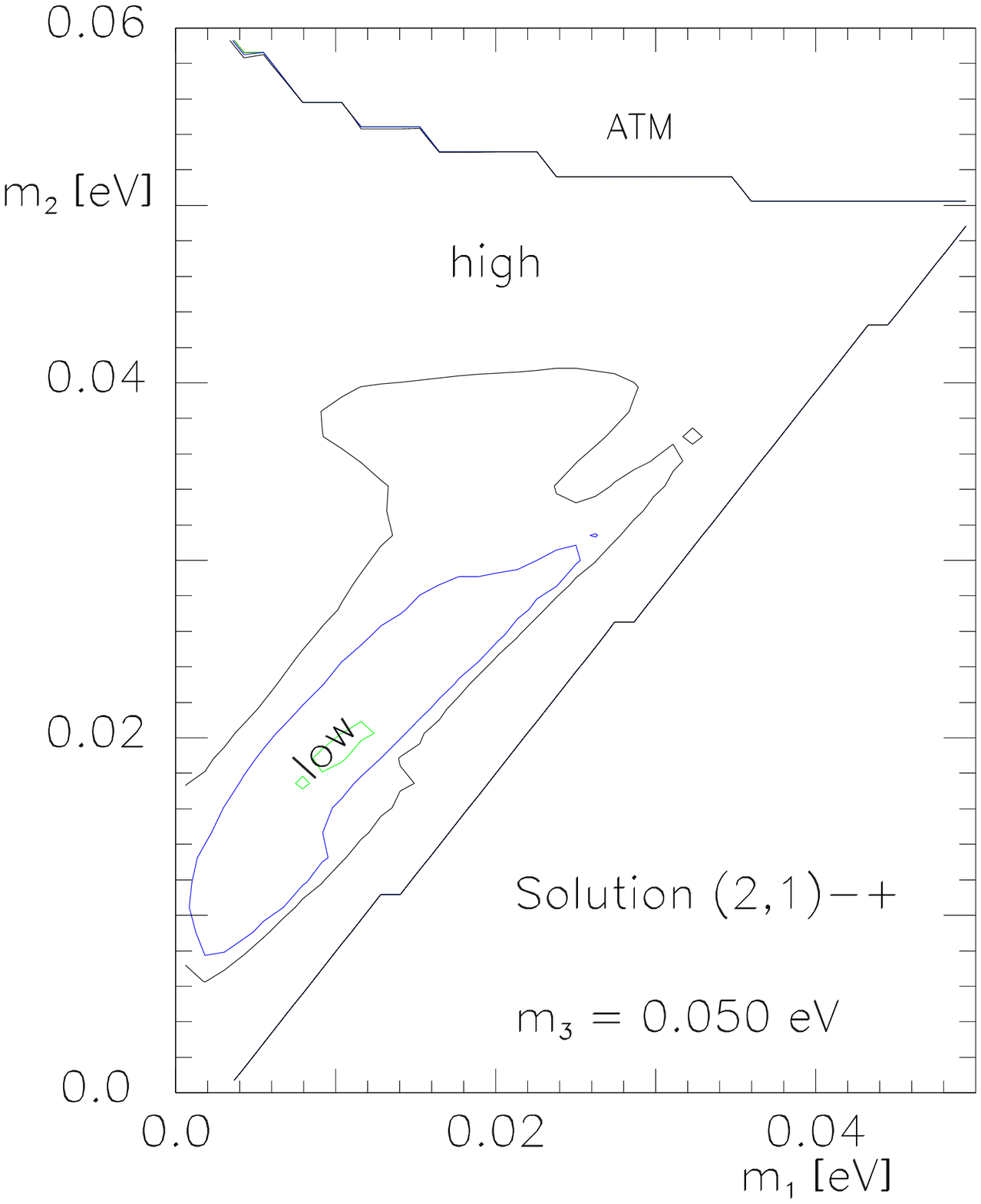}}
\mbox{\epsfysize=7.5cm\epsffile{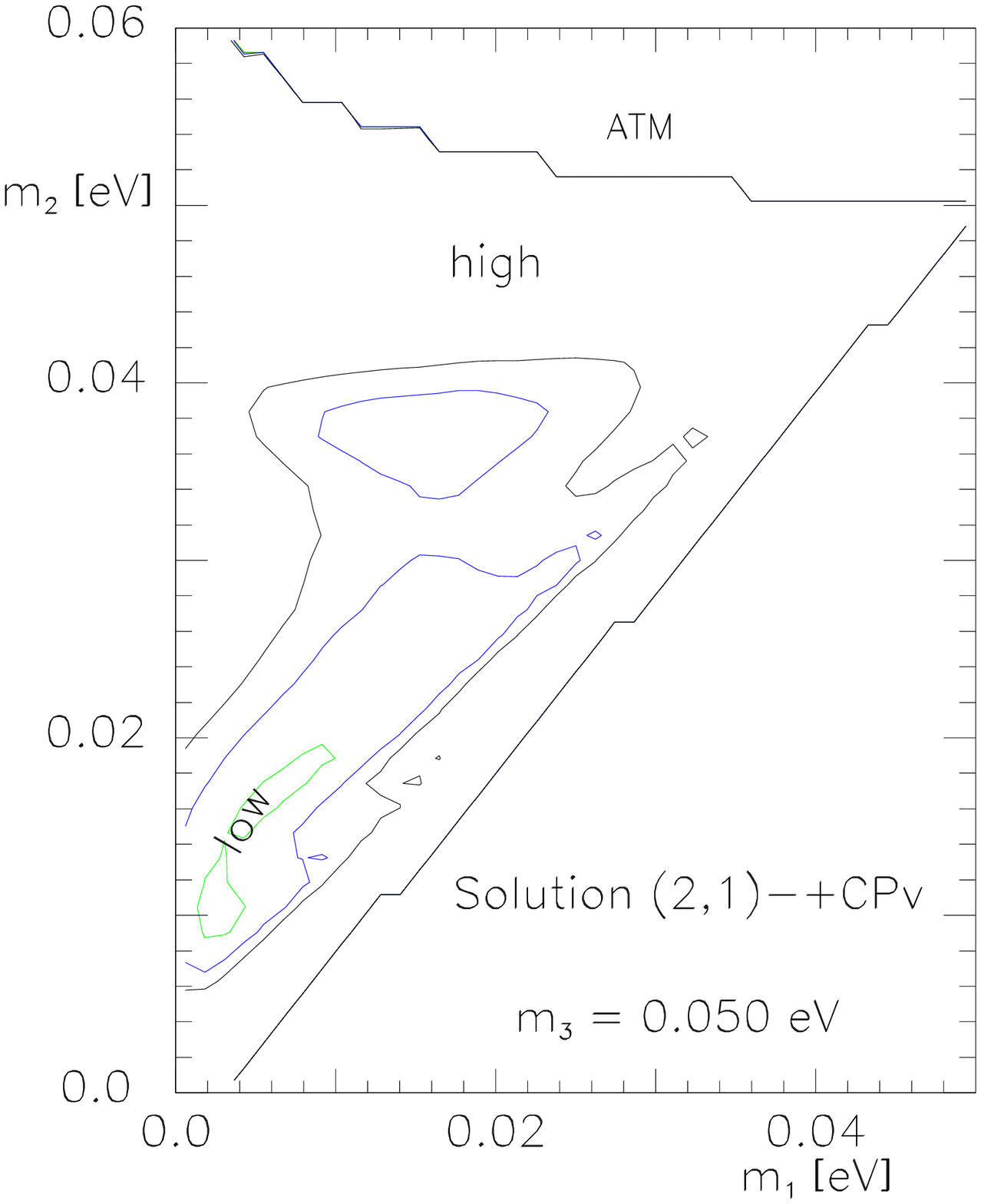}}}
\put(1.,6.0){
\mbox{\epsfysize=7.5cm\epsffile{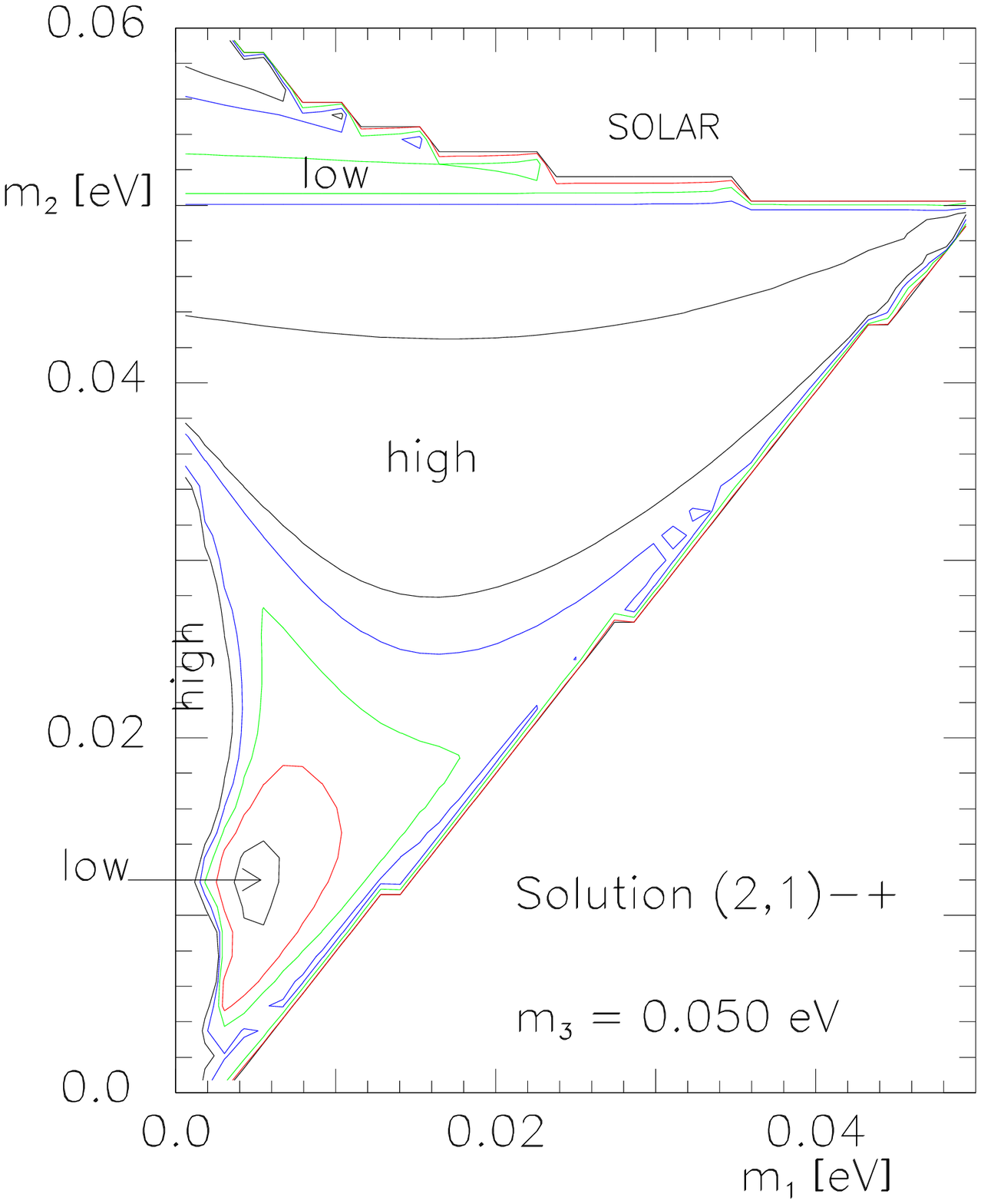}}
\mbox{\epsfysize=7.5cm\epsffile{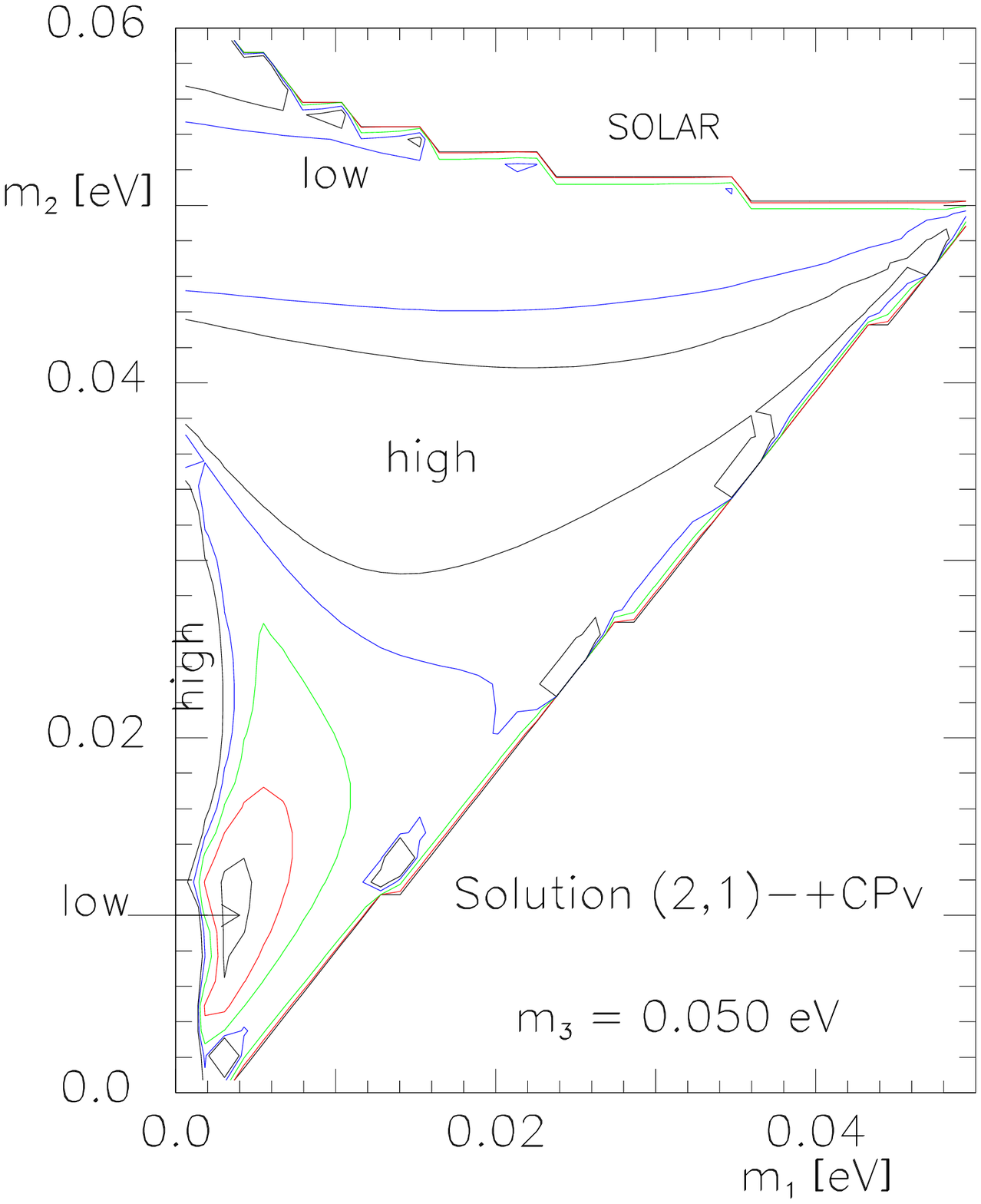}}}
\put(1.,-1.0){
\mbox{\epsfysize=7.5cm\epsffile{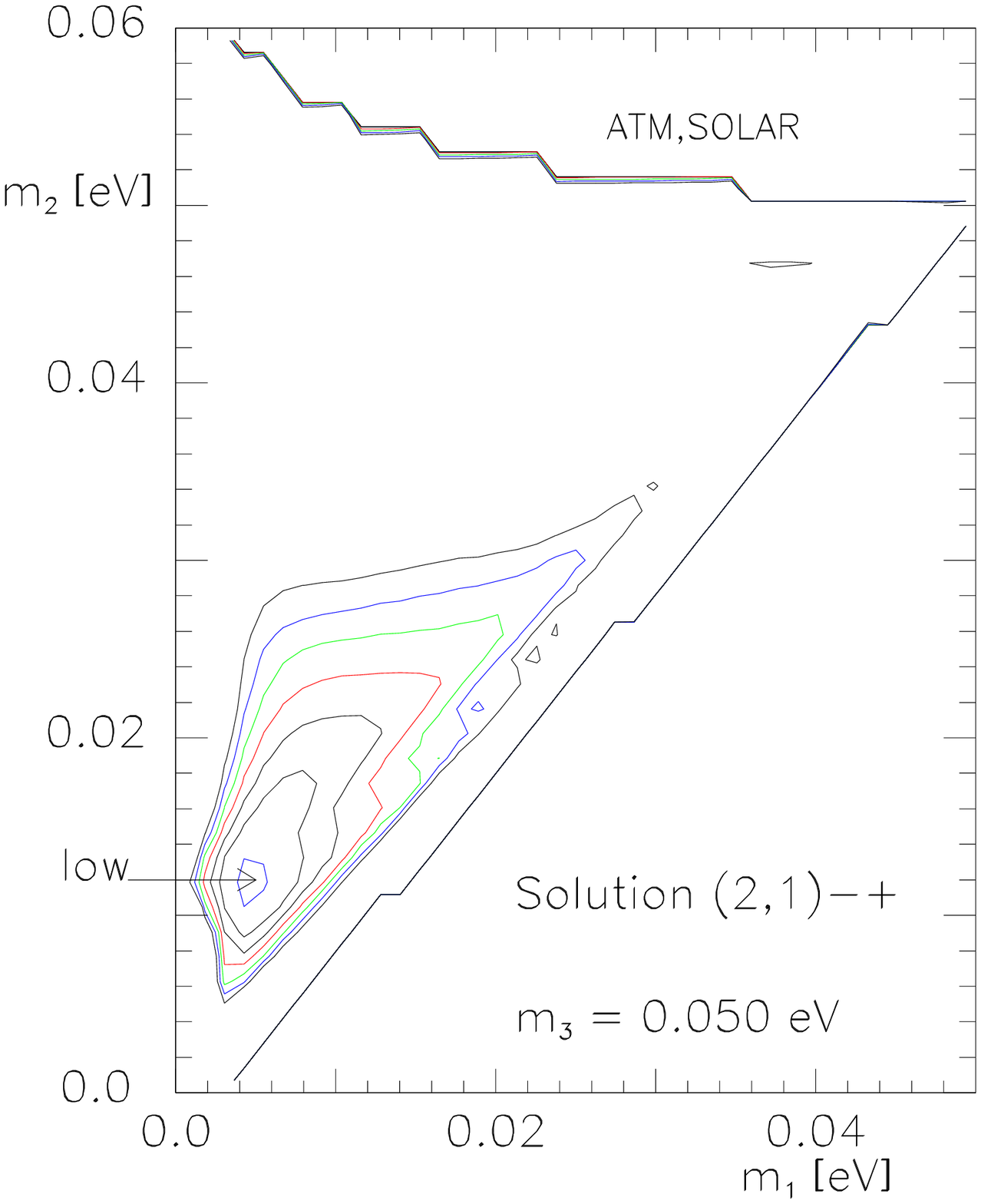}}
\mbox{\epsfysize=7.5cm\epsffile{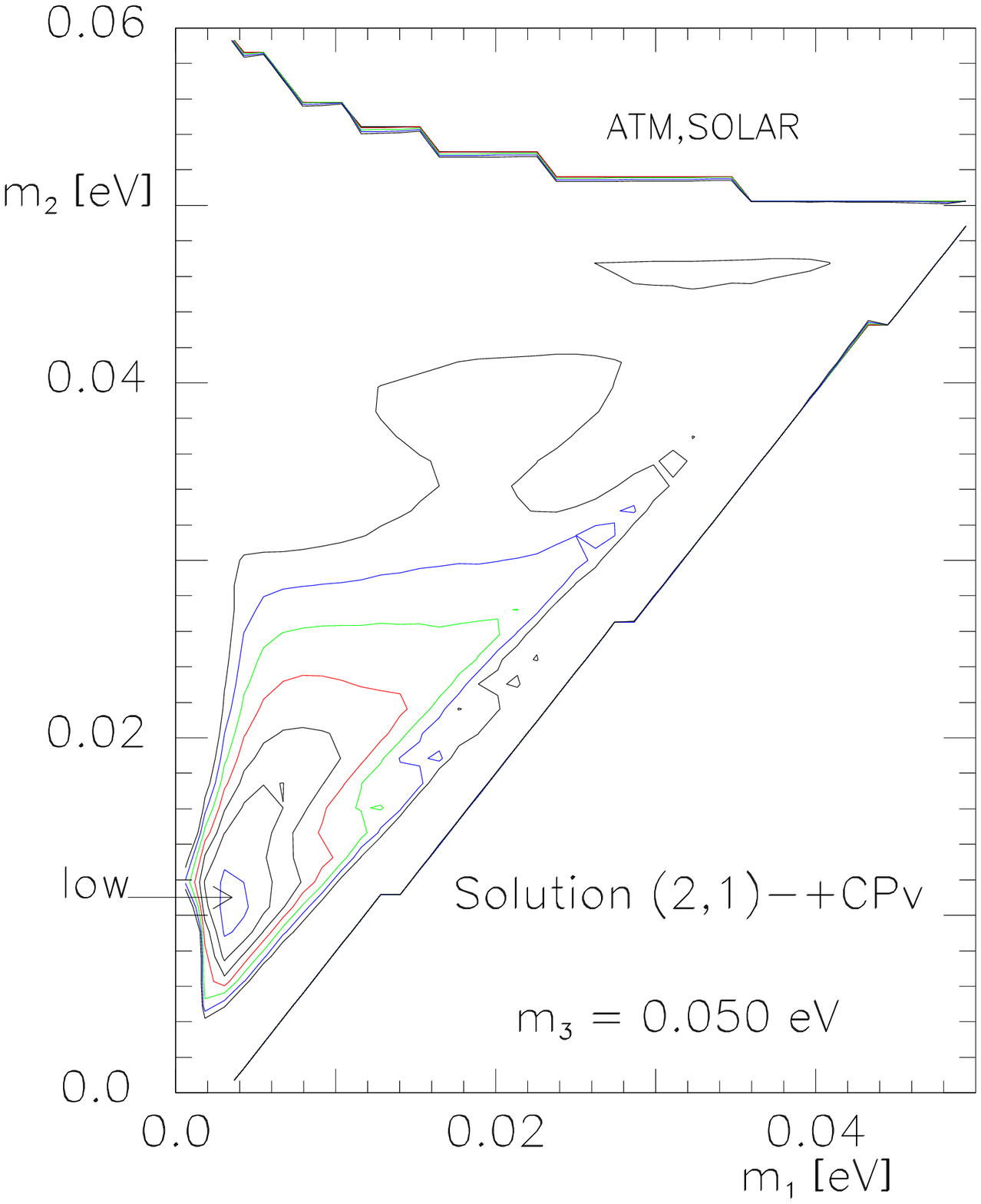}}}
\end{picture}
\vspace*{4mm}
\caption{ Fits to the atmospheric and solar neutrino data (procedure~A) for
the mixing matrix denoted\break ``$(2,1)-+$'' (see Sec.~\ref{sec:5}). 
Contours are given at $\chi^2=5$, 10, 15, 20, 25.  Sum also at 30, 35, 40,
45, 50.  Left panels: $CP$ conservation; right panels: $CP$
non-conservation.}
\end{center}
\end{figure}

The lower panels in Fig.~\ref{Fig:lvls-05} give the corresponding total
$\chi^2=\chi^2_{\text{atm}}+\chi^2_{\text{solar}}$. Since the minimum in
$\chi^2_{\text{solar}}$ is rather well-localized (``steep''), the location
of the overall minimum (for fixed $m_3$) is largely determined by the
minimum in $\chi^2_{\text{solar}}$. However, the extent to which the
minimum in $\chi^2_{\text{atm}}$ overlaps with that of
$\chi^2_{\text{solar}}$ determines how good the fit is.

The cases considered here are $U(2,1)-+$, with $CP$ conservation (left) and
$CP$ non-conservation (right).  Similar scans for the other mixing matrices
(other Solutions, other parities) yield rather similar results.  There is 
little difference between the cases of $CP$ conservation and $CP$
non-conservation.

\section{Results}\label{sec:9}

\subsection{Combining Atmospheric and Solar Data}\label{subsec:9A}

Starting at the minima found from two-dimensional scans as described in
Secs.~\ref{sec:6} and \ref{sec:8}, but now with $m_1$, $m_2$ and $m_3$ all
free, we find total minima for a given solution of the mixing matrix. The
fitting has been performed using two different procedures.  Procedure~A is
the method proposed by Hata and Langacker \cite{Hata:1993rk}, where we
allow for correlations between different data.  Procedure~B is a more
transparent approach, where the data are treated as uncorrelated, and the
SNO data are left out. There are 16 degrees of freedom.  The detailed
numerical results differ somewhat between these two procedures.

\begin{table}[t]
\begin{center} 
\caption{Best fits to the atmospheric and solar neutrino data from
the fitting procedure~A. The number of degrees of freedom is~18,
masses are in eV.}\label{tab:3}
\vspace{.175in}
\renewcommand{\tabcolsep}{1em}
\begin{tabular}{|c|c|c|c|c|c|c|l|} 
\hline 
Solution&$m_1$&$m_2$&$m_3$&$\chi^2_{\text{atm}}$&$\chi^2_{\text{solar}}$
&$\chi^2$&\\
\hline
$(1,1)-+$&0.0049&0.013&0.038& 19.2 & 3.8 & 23.0 &$CP$ cons. \\
\multicolumn{1}{|l|}{same}&0.0036&0.012&0.037& 18.0 & 3.7 & 21.7 
&$CP$ non-cons.
\\
$(1,1)--$&0.0027&0.011&0.038& 17.1 & 3.7 & 20.8 &$CP$ cons. \\
\multicolumn{1}{|l|}{same}&0.0036&0.012&0.038& 17.7 & 3.7 & 21.4 &$CP$
non-cons. \\
$(1,2)++$&0.0047&0.008&0.052& 14.9 & 5.8 & 20.7 &$CP$ cons. \\
\multicolumn{1}{|l|}{same}&0.0030&0.007&0.052& 15.3 & 5.7 & 21.0 &$CP$
non-cons. \\
$(1,2)+-$&0.0022&0.006&0.052& 16.5 & 6.1 & 22.6 &$CP$ cons. \\
\multicolumn{1}{|l|}{same}&0.0030&0.007&0.052& 16.3 & 5.9 & 22.1 &$CP$
non-cons. \\
$(2,1)++$&0.0045&0.008&0.052& 15.1 & 6.1 & 21.2 &$CP$ cons. \\
\multicolumn{1}{|l|}{same}&0.0028&0.006&0.052& 15.4 & 5.3 & 20.8 &$CP$
non-cons. \\
$(2,1)+-$&0.0021&0.006&0.052& 16.7 & 6.0 & 22.7 &$CP$ cons. \\
\multicolumn{1}{|l|}{same}&0.0028&0.007&0.052& 16.6 & 5.3 & 21.9 &$CP$
non-cons. \\ 
\multicolumn{1}{|l|}{same}&0.0029&0.007&0.095& 16.3 & 4.5 & 20.8 &$CP$
non-cons. \\
$(2,1)-+$&0.0045&0.013&0.052& 14.5 & 3.6 & 18.1 &$CP$ cons. \\
\multicolumn{1}{|l|}{same}&0.0035&0.011&0.053& 13.8 & 3.6 & 17.3 &$CP$
non-cons. \\
$(2,1)--$&0.0026&0.010&0.052& 12.5 & 3.8 & 16.3 &$CP$ cons. \\
\multicolumn{1}{|l|}{same}&0.0034&0.011&0.052& 13.1 & 3.7 & 16.8 &$CP$
non-cons. \\
\hline
\end{tabular} 
\end{center}
\end{table}

\begin{table}[ht]
\begin{center}
\caption{Best fits to the atmospheric and solar neutrino data from
the fitting procedure~B.  The number of degrees of freedom
is~16, masses are in eV.}\label{tab:4}
\vspace{.175in}
\renewcommand{\tabcolsep}{1em}
\begin{tabular}{|c|c|c|c|c|c|c|l|}
\hline
Solution&$m_1$&$m_2$&$m_3$&$\chi^2_{\text{atm}}$&$\chi^2_{\text{solar}}$
&$\chi^2$&\\
\hline
$(2,1)-+$&0.0047&0.011&0.054& 16.2 & 0.8 & 17.0 &$CP$ cons. \\
\multicolumn{1}{|l|}{same}&0.0034&0.010&0.054& 15.0 & 0.8 & 15.8 &$CP$
non-cons. \\
$(2,1)--$&0.0025&0.010&0.054& 15.0 & 0.8 & 15.8 &$CP$ cons. \\
\multicolumn{1}{|l|}{same}&0.0035&0.010&0.054& 15.3 & 0.8 & 16.1 &$CP$
non-cons. \\
\hline
\end{tabular}
\end{center}
\end{table}

The best such results from fitting procedure~A are collected in
Table~\ref{tab:3}. Two cases stand out. They are Solution ``$(2,1)--$'' and
``$(2,1)-+$,'' with $\chi^2$ values ranging from 16.3 to 18.1.
Thus, Solution~2 is favored for the neutrino mass matrix,
whereas Solution~1 is favored for the charged lepton mass matrix.
Also, negative ``$b$-parity'' is favored, whereas there is no clear
preference for a particular ``$d$-parity.'' The latter observation is
commented on in Sec.~\ref{sec:10}.

With fitting procedure~B, these two Solutions have $\chi^2$
ranging from 15.8 to 17.0 as can be seen from Table~\ref{tab:4}.

There is a strong clustering of $m_3$ values at $0.052\text{--}0.054$~eV,
for different Solutions and with different ``parities.''
However, there are also a few fits which are not much inferior
with $m_3$ values around $0.04$~eV, and one at $0.095$~eV. 
These have a $\chi^2$ that is higher by about 4 units.

The best-fit mass values may roughly be related to the conventional
\cite{Gonzalez-Garcia:2000sq,Fogli:2001xt}
atmospheric and solar neutrino parameters as
$\Delta m^2_{\text{atm}}\simeq m_3^2-m_2^2=2.6\times10^{-3}~\text{eV}^2$
and $\Delta m^2_{\text{solar}}\simeq m_2^2-m_1^2
=(0.9\text{--}1.5)\times10^{-4}~\text{eV}^2$.

\subsection{Impact of CHOOZ Data}\label{subsec:9B}

The CHOOZ data \cite{Apollonio:1997xe} are known to disfavor models with
``large'' values of $|U_{e3}|$. In our model, large $|U_{e3}|$ typically
require $m_2$ comparable with $m_3$, as can be seen in the example shown in
Fig.~\ref{Fig:U21mp-cont-1}. Thus, we do not expect the minima shown in
Table~\ref{tab:3} to be significantly altered by the inclusion of the CHOOZ
data. This is in fact the case, as shown in Table~\ref{tab:5}.  For the fits of
Table~\ref{tab:3}, the inclusion of the CHOOZ data (14 data points) increases
the total $\chi^2$ by 3--4 units, without changing the best-fit mass values.

\begin{table}[t]
\begin{center}
\caption{Best fits from the fitting procedure~A, 
including the CHOOZ data \cite{Apollonio:1997xe}.
Masses are in eV.}\label{tab:5}
\vspace{.175in}
\renewcommand{\tabcolsep}{.85em}
\begin{tabular}{|c|c|c|c|c|c|c|c|l|}
\hline 
Solution&$m_1$&$m_2$&$m_3$&$\chi^2_{\text{atm}}$
&$\chi^2_{\text{solar}}$
&$\chi^2_{\text{CHOOZ}}$&$\chi^2$&\\
\hline
$(2,1)-+$&0.0044&0.013&0.052& 14.5 & 3.6 & 2.9 & 21.0 &$CP$ cons. \\
\multicolumn{1}{|l|}{same}&0.0034&0.011&0.052& 13.6 & 3.7 & 3.1 & 20.4 &$CP$
non-cons.
\\
$(2,1)--$&0.0026&0.010&0.053& 12.6 & 3.8 & 3.9 & 20.2 &$CP$ cons. \\
\multicolumn{1}{|l|}{same}&0.0034&0.011&0.052& 13.1 & 3.6 & 3.1 & 19.9 &$CP$
non-cons.
\\
\hline
\end{tabular} 
\end{center}
\end{table}

\begin{table}[ht]
\begin{center}
\caption{Mixing matrix elements $U_{\alpha j}$ ($\alpha=e,\mu,\tau$) 
for the best fits, given as (modulus, phase/$\pi$).
Masses are in eV.}\label{tab:6}
\vspace{.175in}
\renewcommand{\tabcolsep}{.65em}
\begin{tabular}{|c|c|c|c|c|c|c|r@{}l|}
\hline 
Solution&$m_1$&$m_2$&$m_3$
&$U_{\alpha1}$&$U_{\alpha2}$&$U_{\alpha3}$
&\multicolumn{2}{c|}{$J_{CP}$}\\
\hline
           & & & &$(0.87,0.00)$&$(0.48,1.00)$&$(0.13,0.00)$&&\\
$(2,1)-+$&0.0044&0.013&0.052&$(0.39,0.00)$&$(0.81,0.00)$&$(0.44,0.00)$
&0.&00\\
          & & & &$(0.31,1.00)$&$(0.33,1.00)$&$(0.89,0.00)$&&\\
\hline
           & & & &$(0.86,-0.49)$&$(0.50,0.47)$&$(0.08,-0.38)$&&\\
$(2,1)-+$&0.0034&0.011&0.052&$(0.43,0.04)$&$(0.79,-0.01)$&$(0.44,0.00)$
&0.&0060\\
          & & & &$(0.28,1.00)$&$(0.34,1.00)$&$(0.90,0.00)$&&\\
\hline
           & & & &$(0.85,0.00)$&$(0.52,0.00)$&$(0.03,1.00)$&&\\
$(2,1)--$&0.0026&0.010&0.053&$(0.46,1.00)$&$(0.77,0.00)$&$(0.44,0.00)$
&0.&00\\
          & & & &$(0.26,0.00)$&$(0.36,1.00)$&$(0.90,0.00)$&&\\
\hline
           & & & &$(0.86,-0.51)$&$(0.51,-0.47)$&$(0.08,0.38)$&&\\
$(2,1)--$&0.0034&0.011&0.052&$(0.43,0.96)$&$(0.79,0.01)$&$(0.43,0.00)$
&$-$0.&0060\\
          & & & &$(0.28,0.00)$&$(0.34,1.00)$&$(0.90,0.00)$&&\\ 
\hline
\end{tabular} 
\end{center}
\end{table}

We show in Table~\ref{tab:6} the relevant mixing elements corresponding to
these best fits.
Since these cases all refer to Solution~1 for the charged leptons,
then to a good approximation we have $U \approx R(\ell)$.
Hence, the cases presented here are pairwise related ($d \to -d$)
by the symmetry of $R(\nu)$, discussed in Sec.~\ref{sec:4}.

In the decoupling approximation (neglecting $U_{e3}$, which ranges
from 0.03 to 0.13), the relevant 
quantities for solar and atmospheric neutrinos are 
$|U_{e1}U_{e2}|\simeq 0.42\text{--}0.44$ and
$|U_{\mu3}U_{\tau3}|\simeq 0.39\text{--}0.40$, respectively,
corresponding to large mixing in both cases.
As shown in Table~\ref{tab:6}, when $CP$ is not conserved, the amount is given
by 
\begin{equation}  \label{Eq:leptons-J}
J_{CP} = \pm6.0 \times10^{-3}
\end{equation}
for neutrinos.  This is much larger than the 
corresponding quantity for quarks as given by Eq.~(\ref{Eq:quarks-J}).

\section{Discussion}\label{sec:10}

Our theoretical description of neutrino oscillations suffers from the fact
that the calculation of the three-flavor MSW effect relies on a somewhat crude
solar model electron density. Nevertheless, it is most rewarding to find that
the determination of the three neutrino masses can be carried out
successfully, i.e., in very reasonable agreement with the experimental data.
We have found excellent fits to the data with
$m_3=(52\text{--}54)\times10^{-3}$~eV, $m_2=(10\text{--}13)\times10^{-3}$~eV,
and $m_1=(2\text{--}5)\times10^{-3}$~eV.

In a regime where 
\begin{equation}  \label{Eq:hierarchy}
m_1\ll m_2\ll m_3, 
\end{equation}
the atmospheric neutrino data
determine $m_3=\mathcal{O}(\sqrt{\Delta m^2_{\text{atm}}})\simeq 0.05$~eV,
and the solar neutrino data determine
$m_2=\mathcal{O}(\sqrt{\Delta m^2_{\text{solar}}})\simeq 0.01$~eV.
The test of the model then lies in (i) reproducing the hierarchy
(\ref{Eq:hierarchy}) and (ii) the determination of $m_1$.

The fact that large mixing is required by the atmospheric as well as the solar
neutrino data essentially forces the model into a region of parameter space
where there is a strong hierarchy. To some extent, this can be read off from
Figs.~\ref{Fig:Rnu-cont-1}--\ref{Fig:U21mp-cont-1}.

Another issue is to what extent the data can distinguish the different
discrete parameters of the model, Solution~1 \textit{vs.}\ Solution~2,
as well as the $b$ and $d$ ``parities.''
The data favor Solution~2 for the neutrino mixing and Solution~1 for
the charged leptons, both by a margin of 4.4 units of $\chi^2$.

The best fits have been found for charged lepton Solution~1.
Since the charged lepton masses are strongly hierarchical, the rotation
matrix corresponding to Solution~1 is very close to the unit matrix.
Thus, the overall neutrino mixing matrix $U$ is rather close
to $R(\nu)$. This explains why the mass values obtained are close
to those presented earlier \cite{Osland:2000bh}.

It is of some interest to compare in more detail with the case
\begin{equation}
U=R(\nu).
\end{equation}
For the two best fits, Solutions $(2)-+$ and $(2)--$, we find
$\chi^2=17.2$ and 16.7, respectively, with $m_3=0.057$ and $0.052$~eV. The
former solution is the best fit of \cite{Osland:2000bh}.

Actually, in the limit of no mixing in the charged-lepton sector, $R(\ell)=1$,
when $U=R(\nu)$, we see that the atmospheric transition probability
(\ref{Eq:trans-prob}) is invariant under $b\to-b$, as well as under $d\to-d$
or $d\to id$.  Similarly, for the MSW equation (\ref{Eq:M-squared}), the sign
change $b\to-b$ can be compensated for by a wave function sign change,
$\phi_3\to-\phi_3$, and the sign change $d\to-d$ (or $d\to id$) can be
compensated for by sign (or phase) changes of $\phi_2$ and $\phi_3$.  Thus,
these are exact symmetries of $|\phi_1|^2$ in this limit of $R(\ell)=1$.  They
are therefore approximate symmetries for Solution~1 for the charged-lepton
sector.

It is also interesting to compare with the rather different model
proposed by \cite{Koide:2002nc}. The masses they find are $m_3=0.0506$~eV,
$m_2=(7.46\text{--}7.48)\times10^{-3}$~eV, and
$m_1=(2.39\text{--}2.43)\times10^{-3}$~eV. As mentioned above,
the fact that $m_3$ and $m_2$ are rather similar to the values we find
is unavoidable within a hierarchical fit.

We add a comment on $CP$ non-conservation in the lepton sector.  In the 
work of Lehmann \textit{et al.}~\cite{Lehmann:1995br}, given the other
parameters, $CP$ non-conservation is maximal.  For this reason, when
$\epsilon = i$, the  lepton $CP$ non-conservation is also maximal in the same
sense.  In  other words, the lepton $CP$ non-conservation is either zero 
($\epsilon = 1$) or maximal ($\epsilon = i$).  Thus there are only two 
cases instead of a continuum.  

The absolute values of the elements of the mixing matrices are somewhat
different in these two cases.  These differences are not sufficiently large to
decide experimentally whether $CP$ is conserved or not in the lepton sector,
since both give comparable fits, as shown in Table~\ref{tab:5}.  Note that the
$\chi^2$ is slightly smaller (by 0.3--0.6) for the case of $CP$
non-conservation. That the Jarlskog determinant (\ref{Eq:leptons-J}) for
leptons is significantly larger than that of (\ref{Eq:quarks-J}) for quarks is
related to the fact that the ratios of neutrino masses obtained here for
leptons are larger than those for quarks (for a more general discussion, see
\cite{Giunti:2002pp}).
\bigskip

\begin{acknowledgments}
\par\noindent
It is a pleasure to thank Geir Vigdel for very useful discussions.
One of us (TTW) wishes to thank the Theory Division of CERN for its kind
hospitality.
\hfil\break
\indent 
This work was supported in part by the Fund for Scientific Research,
Flanders-Belgium and by the Interuniversity Attraction Poles, Belgium 
(P5-11-35);
in part by the Research Council of Norway; and in part by the United
States Department of Energy under Grant No.\ DE-FG02-84ER40158.
\end{acknowledgments}

\appendix

\section{}\label{app:A}

We discuss in this appendix some elementary properties of the 
mass matrix of Ref.~\cite{Lehmann:1995br}
\begin{equation}
M=\left(\begin{matrix}
0    & id & 0 \\
-id & c & b \\
0    & b & a
\end{matrix}\right),
\end{equation}
with $b^2=8c^2$. From (\ref{Eq:diagonal-nu}), this $M$ is diagonalized by
\begin{equation}
\label{Eq:def-R}
M=\text{diag}(i,1,1)\,R\,\text{diag}(m_1,-m_2,m_3)\,
R^{\text{T}}\,\text{diag}(-i,1,1).
\end{equation}
The rotation matrix $R$ is discussed in Sec~\ref{sec:4}.
The case of the mass matrix without the factors of $i$ and $-i$
is entirely similar.
The relations between the elements of $M$ and the masses have been given
in Eq.~(\ref{Eq:quarks:m1m2m3}).

Consider first the triangular region (\ref{Eq:quark-mass-inequality})
studied in Ref.~\cite{Lehmann:1995br}.
In this region, the $S_1$, $S_2$ and $S_3$ of (\ref{Eq:quarks:m1m2m3})
satisfy
\begin{equation}
\label{Eq:ineq-S1S2S3}
S_3-S_1S_2=(m_2-m_1)(m_3-m_2)(m_3+m_1) > 0.
\end{equation}
Also from (\ref{Eq:quarks:m1m2m3}), the 33 entry of $M$, namely $a$,
satisfies the cubic equation (\ref{Eq:cubic-a}), and the possible signs 
of $a$ have been discussed immediately thereafter.
Since $c=S_1-a$, the corresponding cubic equation for $c$ is
\begin{equation}
\label{Eq:cubic-c}
9c^3-10S_1 c^2+(S_1^2+S_2)c+(S_3-S_1S_2)=0.
\end{equation}

When (\ref{Eq:cubic-a}) has three real solutions, so does (\ref{Eq:cubic-c}).
Furthermore, by the inequality (\ref{Eq:ineq-S1S2S3}), two of the real
solutions must be positive, while the third one is negative.
A comparison with what is known about (\ref{Eq:cubic-a}) shows that,
in the triangular region (\ref{Eq:quark-mass-inequality}),
\begin{alignat}{2}
\label{Eq:c-signs-triangle}
&c<0 &\quad &\text{for Solution 1,} \nonumber \\
&c>0 &\quad &\text{for Solution 2.}
\end{alignat}

When $m_2\to m_1$ or $m_3\to m_2$, it follows from (\ref{Eq:ineq-S1S2S3})
that $S_3-S_1S_2=0$. In either limit, (\ref{Eq:cubic-c}) becomes a
quadratic equation when $c\ne0$,
\begin{equation}
\label{Eq:quadratic-c}
9c^2-10S_1 c+(S_1^2+S_2)=0.
\end{equation}
In the limit $m_2 \to m_1$, this has two positive roots. 
In the limit $m_2 \to m_3$, this has one positive and one negative root.
Therefore, for Solution~1, $c$ remains negative
when $m_2 \to m_1$ or $m_2 \to m_3$.
On the other hand, for Solution~2,
$c$ remains positive when $m_2 \to m_1$, whereas
\begin{equation}
c\to 0
\end{equation}
when $m_2 \to m_3$.\goodbreak

{}From the discussion after (\ref{Eq:cubic-a}), the interesting case
is where this cubic equation has three real roots.
This holds not only in the triangular region (\ref{Eq:quark-mass-inequality})
but in a larger region.
The boundary of this larger region is given by the straight line $m_1=0$
together with curves obtained by setting the discriminant
of the cubic equation (\ref{Eq:cubic-a}) to zero:
\begin{equation}
(8S_1^2+S_2)^2(S_1^2-36S_2)-2187S_3^2
+2380S_1^3S_3+2754S_1S_2S_3=0.
\end{equation}
The resulting region is only slightly larger than the triangle
(\ref{Eq:quark-mass-inequality}), and has been discussed in Sec.~\ref{sec:3}.
Let the region shown in Fig.~\ref{Fig:disc} be called $\mathcal{R}$, 
while $\mathcal{R}_0$
denotes the triangle (\ref{Eq:quark-mass-inequality}); then
$\mathcal{R}$ consists of $\mathcal{R}_0$ together with $\mathcal{R}_1$ (the
small region where $m_1>m_2$) and $\mathcal{R}_2$ (the region where
$m_2>m_3$). The extension of (\ref{Eq:c-signs-triangle}) to
$\mathcal{R}_1$ and $\mathcal{R}_2$ gives simply
\begin{alignat}{2}
&c<0 &\quad &\text{for Solution 1 in all of }\mathcal{R} \nonumber \\
&c>0 &\quad &\text{for Solution 2 in }\mathcal{R}_0
\text{ and }\mathcal{R}_1 \nonumber \\
&c<0 &\quad &\text{for Solution 2 in }\mathcal{R}_2.
\end{alignat}

The next property to be discussed is the behavior of the $m$'s
when both $c$ and $d$ are small compared with $a$:
\begin{equation}
\label{Eq:small-b}
|c| \ll a \quad \text{and} \quad |d| \ll a,
\end{equation}
the relative magnitude of $c$ and $d$ being arbitrary.
In general, by Eq.~(\ref{Eq:quarks:m1m2m3}), the masses $m_1$,
$-m_2$ and $m_3$ satisfy the cubic equation
\begin{equation}
\lambda^3-(a+c)\lambda^2+(ac-8c^2-d^2)\lambda+ad^2=0.
\end{equation}
In the limiting case (\ref{Eq:small-b}), one of the solutions is
\begin{equation}
\lambda=m_3\sim a,
\end{equation}
while the other two solutions are both small.
These two small solutions, $m_1$ and $-m_2$, are determined
approximately by the quadratic equation
\begin{equation}
\label{Eq:quadratic-a}
-a\lambda^2+(ac-d^2)\lambda+ad^2=0.
\end{equation}
The solutions of (\ref{Eq:quadratic-a}) are
\begin{align}
\lambda&=\frac{1}{2a}[ac-d^2\pm\sqrt{(ac-d^2)^2+4a^2d^2}] \nonumber \\
&\sim\textstyle\frac{1}{2}[c\pm\sqrt{c^2+4d^2}],
\end{align}
or
\begin{align}
\label{Eq:small-m1,m2}
m_1&=\textstyle\frac{1}{2}[c+\sqrt{c^2+4d^2}], \nonumber \\
m_2&=\textstyle\frac{1}{2}[-c+\sqrt{c^2+4d^2}].
\end{align}
Note that these approximate solutions are applicable only to
Solution~1, because Eq.~(\ref{Eq:limit:m1=m2=0}) implies that,
for Solution~2, $c$ cannot satisfy the inequality (\ref{Eq:small-b}).
Eq.~(\ref{Eq:small-m1,m2}) shows directly that $c<0$ for Solution~1
in the triangular region (\ref{Eq:quark-mass-inequality}).

\section{}\label{app:B}

For completeness we give here asymptotic formulas for the ${}_3F_1$
required for reconstructing the ${}_2F_2$.
These asymptotic formulas turn out to be quite useful and in particular
are accurate for the regions of the fits discussed in Sec.~\ref{sec:8}.
For the wave functions of the electron neutrino, $\psi_1$ of
Eq.~(\ref{Eq:Schr-4}), the two relevant ${}_3F_1$ are
(cf.\ \cite{Osland:1999et})
\begin{align}
\label{Eq:first-F31}
&{}_3F_1\left[\left.
\begin{matrix}
1+i\xi, & 1-i\eta, & 1+i\zeta \\
1+i\zeta'
\end{matrix}
\right|\frac{i}{y} \right] \nonumber \\[4pt]
&\qquad\mbox{}=\frac{1}{(2\pi)^2}\left(e^{i\pi/2}y\right)^{1+i\xi}\,
e^{\pi\zeta'}\;
\frac{\Gamma(1+i\zeta')\Gamma(i\eta)\Gamma(1-i\eta-i\zeta')}
{\Gamma(1+i\xi)}\, I, \\[4pt]
\label{Eq:second-F31}
&{}_3F_1\left[\left.
\begin{matrix}
1+i\bar\xi, & 1-i\bar\eta, & 1-i\bar\zeta \\
1-i\bar\zeta'
\end{matrix}
\right|\frac{i}{y} \right] \nonumber \\[4pt]
&\qquad\mbox{}=\left(e^{i\pi/2}y\right)^{1+i\bar\xi}\,
\frac{\Gamma(1-i\zeta')}
{\Gamma(1+i\bar\xi)\Gamma(1-i\bar\eta)\Gamma(i\bar\eta-i\zeta')}\, \bar I,
\end{align}
where
\begin{eqnarray}
\label{Eq:int-I}
I&=&\int_0^\infty dt \int_{\mathcal{P}} ds\,
(1+s+st)^{-1}\, \exp\{-i[ty-\xi\ln t+\eta\ln s-(\zeta-\zeta')\ln(1+s)
\nonumber \\
&&\hspace*{50mm}\mbox{}
+\zeta\ln(1+s+st)]\}, \\
\label{Eq:int-bar-I}
\bar I&=&\int_0^\infty dt \int_0^\infty ds\,
(1+s+st)^{-1}\, \exp\{-i[ty-\bar\xi\ln t+\bar\eta\ln s
+(\bar\zeta-\zeta')\ln(1+s)
\nonumber \\
&&\hspace*{50mm}\mbox{}
-\bar\zeta\ln(1+s+st)]\},
\end{eqnarray}
\begin{alignat}{4}
\label{Eq:xi-eta-etc}
\xi&=\mu_2-\omega_2, \quad \eta&\mbox{}=\omega_2-\mu_1, \quad
\zeta&=\mu_3-\omega_2, \quad \zeta'&\mbox{}=\omega_3-\omega_2, \nonumber
\\
\bar\xi&=\mu_3-\omega_3, \quad \bar\eta&=\omega_3-\mu_2, \quad
\bar\zeta&=\omega_3-\mu_1, 
\end{alignat}
and
\begin{equation}
y=e^{-u}.
\end{equation}
Note that all these quantities in Eq.~(\ref{Eq:xi-eta-etc}) are positive 
because
\begin{equation}
\label{Eq:interlace}
\mu_1 < \omega_2 < \mu_2 < \omega_3 < \mu_3.
\end{equation}

The double integrals for $I$ and $\bar I$ of Eqs.~(\ref{Eq:int-I})
and (\ref{Eq:int-bar-I}) can be carried out approximately
using the method of stationary phase, and the resulting
asymptotic formulas are
\begin{eqnarray}
I&\sim& 2\pi\, \sigma_0^{-1/2}
\biggl[-\frac{\xi(\zeta-\zeta')(\sigma_0-1-t_0)^2}{(\sigma_0-1)^2 t_0^2}
+\frac{\zeta(\zeta-\zeta')}{(\sigma_0-1)^2}
+\frac{(1+t_0)\xi\zeta}{t_0^2}
+\frac{\zeta^2}{\sigma_0-1-t_0}\biggr]^{-1/2} \nonumber \\[4pt]
&&\mbox{}\times \exp\{-i[t_0y-\xi\ln t_0-(\eta+\zeta')\ln\sigma_0
-(\zeta-\zeta')\ln(\sigma_0-1)
\nonumber \\
&& \hspace*{30mm}\mbox{}
+\zeta\ln(\sigma_0-1-t_0)]\}\;,
\end{eqnarray}
and
\begin{eqnarray}
\bar I&\sim& 2\pi\, \bar\sigma_0^{-1/2}
\biggl[-\frac{\bar\xi(\bar\zeta-\zeta')(\bar\sigma_0+1+\bar t_0)^2}
{(\bar\sigma_0+1)^2 \bar t_0^2}
-\frac{\bar\zeta(\bar\zeta-\zeta')}{(\bar\sigma_0+1)^2}
+\frac{(1+\bar t_0)\bar\xi\bar\zeta}{\bar t_0^2}
+\frac{\bar\zeta^2}{\bar\sigma_0+1+\bar t_0}\biggr]^{-1/2} \nonumber \\[4pt]
&&\mbox{}\times \exp\{-i[\bar t_0y-\bar\xi\ln \bar t_0
-(\bar\eta-\zeta')\ln\bar\sigma_0
+(\bar\zeta-\zeta')\ln(\bar\sigma_0+1)
\nonumber \\
&& \hspace*{30mm}\mbox{}
-\bar\zeta\ln(\bar\sigma_0+1+\bar t_0)]\}\;,
\end{eqnarray}
where $\sigma_0$ and $t_0$ are determined by
\begin{eqnarray}
y-\frac{\xi}{t_0}-\frac{\zeta}{\sigma_0-1-t_0}
&=&0, \nonumber \\
\eta+\frac{\zeta-\zeta'}{\sigma_0-1}-\frac{(1+t_0)\zeta}{\sigma_0-1-t_0}
&=&0,
\end{eqnarray}
while $\bar\sigma_0$ and $\bar t_0$ are determined by
\begin{eqnarray}
y-\frac{\bar\xi}{\bar t_0}-\frac{\bar\zeta}{\bar\sigma_0+1+\bar t_0}
&=&0, \nonumber \\
\bar\eta+\frac{\bar\zeta-\zeta'}{\bar\sigma_0+1}
-\frac{(1+\bar t_0)\bar\zeta}{\bar\sigma_0+1+\bar t_0}
&=&0.
\end{eqnarray}
Note that the $\sigma_0$ here is equal to $-1/s_0$, where $s_0$
is defined in the Appendix of \cite{Osland:1999et}.

\section{}\label{app:C}

We shall provide here some further details, beyond what
was given in \cite{Osland:1999et}, on the reconstruction
of $\psi^{(1)}$ and $\psi^{(2)}$ from the ${}_3F_1$'s.

We introduce subscripts $i$ and $k$ to label the cases (5.3)--(5.5) 
and (5.12)--(5.13), respectively, of \cite{Osland:1999et}.
Then, Eqs.~(5.12)--(5.13) of \cite{Osland:1999et} may be rewritten as
\begin{equation}
\label{Eq:f-hat-simple}
\hat f_i^{(k)}(z)
=G_{ik}\left[Y_{ik1}\,f_i^{(1)}(z)+Y_{ik2}\,f_i^{(2)}(z)
            +Y_{ik3}\,f_i^{(3)}(z)\right],
\end{equation}
where
\begin{equation}
G_{i1}=\frac{\pi\Gamma(1-\alpha_2+\alpha_1)}
{\Gamma(-\beta_1+\alpha_1)\Gamma(-\beta_2+\alpha_1)\Gamma(-\beta_3+\alpha_1)
}
\end{equation}
and
\begin{equation}
Y_{i11}=\frac{\Gamma(-\beta_2+\beta_1)\Gamma(-\beta_3+\beta_1)}
{\Gamma(1-\alpha_1+\beta_1)\Gamma(1-\alpha_2+\beta_1)}
\,\frac{1}{\sin\pi(\alpha_1-\beta_1)},
\end{equation}
etc., with $\alpha_i$ and $\beta_j$ given by Eqs.~(5.2)--(5.5)
of \cite{Osland:1999et}.
Furthermore,
\begin{equation}
 \left\{G_{i2},Y_{i2j}\right\}
=\left\{G_{i1},Y_{i1j}\right\}_{\alpha_1\leftrightarrow\alpha_2}.
\end{equation}
Note that $Y_{i2j}=Y_{i1j}$ apart from the change of argument of
$\sin[\pi(\alpha_k-\beta_j)]$.

Define now
\begin{equation}
f^{(j)}(z)=e^{-\pi\mu_j/2}\,\bar f^{(j)}(z), \quad
\hat f^{(k)}(z)=e^{-i(\pi/2)\alpha_k}\,\tilde f^{(k)}(z). 
\end{equation}
Then, Eq.~(\ref{Eq:f-hat-simple}) can be rewritten as
\begin{equation}
\tilde f_i^{(k)}(z)
=B_{ik1}\bar f_i^{(1)}(z)+B_{ik2}\bar f_i^{(2)}(z)+B_{ik3}\bar f_i^{(3)}(z)
\end{equation}
with
\begin{equation}
B_{ikj}=G_{ik}\,Y_{ikj}\,e^{(\pi/2)[i\alpha_k-\mu_j]}.
\end{equation}

For each $i$, we have the following two equations:
\begin{eqnarray}
B_{i11}\bar f_i^{(1)}(z)+B_{i12}\bar f_i^{(2)}(z)
=\left[\tilde f_i^{(1)}(z)-B_{i13}\bar f_i^{(3)}(z)\right], \nonumber \\
B_{i21}\bar f_i^{(1)}(z)+B_{i22}\bar f_i^{(2)}(z)
=\left[\tilde f_i^{(2)}(z)-B_{i23}\bar f_i^{(3)}(z)\right].
\end{eqnarray}
Here, $\tilde f_i^{(1)}(z)$ and $\tilde f_i^{(2)}(z)$ are given in terms
of
${}_3F_1$'s, whereas the $\bar f_i^{(3)}(z)$ are given in terms of
${}_2F_2$'s.
These are then solved for $\bar f_i^{(1)}(z)$ and $\bar f_i^{(2)}(z)$,
from which the ${}_2F_2$ of Eq.~(5.7) are obtained.

To obtain the physical neutrino wave functions, one has to rotate
back to the $\phi_i$ of Eq.~(\ref{Eq:Schr-1}).



\begin{thebibliography}{99}

\bibitem{Lehmann:1995br}
H.~Lehmann, C.~Newton and T.~T.~Wu, 
Phys.\ Lett.\ B \textbf{384}, 249 (1996).

\bibitem{Cabibbo:yz}
N.~Cabibbo, 
Phys.\ Rev.\ Lett.\  \textbf{10}, 531 (1963);
M.~Kobayashi and T.~Maskawa,
Prog.\ Theor.\ Phys.\  \textbf{49}, 652 (1973).

\bibitem{Harari:1978yi}
H.~Harari, H.~Haut and J.~Weyers,
Phys.\ Lett.\ B \textbf{78}, 459 (1978);
P.~Kaus and S.~Meshkov,
Mod.\ Phys.\ Lett.\ A \textbf{3}, 1251 (1988)
[Erratum-ibid.\ A \textbf{4}, 603 (1989)];
M.~Tanimoto,
Phys.\ Rev.\ D \textbf{41}, 1586 (1990);
M.~Fukugita, M.~Tanimoto and T.~Yanagida,
Prog.\ Theor.\ Phys.\  \textbf{89}, 263 (1993);
M.~Fukugita, M.~Tanimoto and T.~Yanagida,
Phys.\ Rev.\ D \textbf{57}, 4429 (1998)
[arXiv:hep-ph/9709388];
Y.~Nomura and T.~Yanagida,
Phys.\ Rev.\ D \textbf{59}, 017303 (1999)
[arXiv:hep-ph/9807325].

\bibitem{Fritzsch:1995dj}
H.~Fritzsch and Z.~Z.~Xing,
Phys.\ Lett.\ B \textbf{372}, 265 (1996)
[arXiv:hep-ph/9509389];
Phys.\ Rev.\ D \textbf{57}, 594 (1998)
[arXiv:hep-ph/9708366].

\bibitem{Altarelli:gu}
G.~Altarelli and F.~Feruglio,
Phys.\ Rept.\  \textbf{320}, 295 (1999);
G.~Altarelli and F.~Feruglio,
arXiv:hep-ph/0206077.

\bibitem{Roberts:2001zy}
R.~G.~Roberts, A.~Romanino, G.~G.~Ross and L.~Velasco-Sevilla,
Nucl.\ Phys.\ B \textbf{615}, 358 (2001)
[arXiv:hep-ph/0104088].

\bibitem{Fukuda:1998tw}
Y.~Fukuda \textit{et al.}  [Super-Kamiokande Collaboration],
Phys.\ Lett.\ B \textbf{433}, 9 (1998)
[arXiv:hep-ex/9803006];
Phys.\ Lett.\ B \textbf{436}, 33 (1998)
[arXiv:hep-ex/9805006];
Phys.\ Rev.\ Lett.\  \textbf{81}, 1562 (1998)
[arXiv:hep-ex/9807003].

\bibitem{Fukuda:1998fd}
Y.~Fukuda \textit{et al.}  [Super-Kamiokande Collaboration],
Phys.\ Rev.\ Lett.\  \textbf{81}, 1158 (1998)
[Erratum-ibid.\  \textbf{81}, 4279 (1998)]
[arXiv:hep-ex/9805021];
S.~Fukuda \textit{et al.}  [Super-Kamiokande Collaboration],
Phys.\ Rev.\ Lett.\  \textbf{86}, 5651 (2001)
[arXiv:hep-ex/0103032].

\bibitem{Glashow:tr}
S.~L.~Glashow,
Nucl.\ Phys.\  \textbf{22}, 579 (1961);
S.~Weinberg,
Phys.\ Rev.\ Lett.\  \textbf{19}, 1264 (1967);
A.\ Salam, in \textit{Elementary Particle Theory: Relativistic Groups and 
Analyticity (Nobel Symposium No. 8)}, edited by N. Svartholm (Almqvist and 
Wiksell, Stockholm, 1968), p. 367;
S.~L.~Glashow, J.~Iliopoulos and L.~Maiani,
Phys.\ Rev.\ D \textbf{2}, 1285 (1970).

\bibitem{review}
For recent reviews, see
E.~K.~Akhmedov,
Lectures given at ICTP Summer School in Particle Physics, Trieste, Italy, 
7 June -- 9 July 1999, 
arXiv:hep-ph/0001264;
W.~C.~Haxton,
to appear in `Current Aspects of Neutrino Physics,' 
ed.\ David Caldwell (Springer-Verlag),
arXiv:nucl-th/0004052;
M.~C.~Gonzalez-Garcia and Y.~Nir,
arXiv:hep-ph/0202058.

\bibitem{King:1999mb}
S.~F.~King,
Nucl.\ Phys.\ B \textbf{576}, 85 (2000)
[arXiv:hep-ph/9912492].

\bibitem{Bilenky:1980cx}
S.~M.~Bilenky, J.~Hosek and S.~T.~Petcov,
Phys.\ Lett.\ B \textbf{94}, 495 (1980).

\bibitem{Cleveland:1994er}
B.~T.~Cleveland \textit{et al.},
Nucl.\ Phys.\ Proc.\ Suppl.\  \textbf{38}, 47 (1995);
Astrophys.\ J.\  \textbf{496}, 505 (1998).

\bibitem{Abdurashitov:1999bv}
J.~N.~Abdurashitov \textit{et al.}  [SAGE Collaboration],
Phys.\ Rev.\ Lett.\  \textbf{83}, 4686 (1999)
[arXiv:astro-ph/9907131];
Phys.\ Rev.\ C \textbf{60}, 055801 (1999)
[arXiv:astro-ph/9907113].

\bibitem{Hampel:1998xg}
W.~Hampel \textit{et al.}  [GALLEX Collaboration],
Phys.\ Lett.\ B \textbf{447}, 127 (1999).

\bibitem{Ahmad:2001an}
Q.~R.~Ahmad \textit{et al.}  [SNO Collaboration],
Phys.\ Rev.\ Lett.\  \textbf{87}, 071301 (2001)
[arXiv:nucl-ex/0106015].

\bibitem{Ahmad:2002jz}
Q.~R.~Ahmad \textit{et al.}  [SNO Collaboration],
Phys.\ Rev.\ Lett.\  \textbf{89}, 011301 (2002)
[arXiv:nucl-ex/0204008].

\bibitem{Osland:2000bh}
P.~Osland and T.~T.~Wu,
preprint, CERN-TH/99-285, 
arXiv:hep-ph/0006185;
Phys.\ Scripta \textbf{T93}, 37 (2001)
[arXiv:hep-ph/0103281].

\bibitem{Jarlskog:1985ht}
C.~Jarlskog,
Phys.\ Rev.\ Lett.\  \textbf{55}, 1039 (1985).

\bibitem{Harrison:1998yr}
P.~F.~Harrison and H.~R.~Quinn  [BABAR Collaboration],
SLAC-R-0504
\textit{Papers from Workshop on Physics at an Asymmetric B Factory 
(BaBar Collaboration Meeting), Rome, Italy, 11-14 Nov 1996, Princeton, NJ, 
17-20 Mar 1997, Orsay, France, 16-19 Jun 1997 and 
Pasadena, CA, 22-24 Sep 1997}.

\bibitem{Christenson:fg}
J.~H.~Christenson, J.~W.~Cronin, V.~L.~Fitch and R.~Turlay,
Phys.\ Rev.\ Lett.\  \textbf{13}, 138 (1964).

\bibitem{Wu:qx}
T.~T.~Wu and C.~N.~Yang,
Phys.\ Rev.\ Lett.\  \textbf{13}, 380 (1964).

\bibitem{Wolfenstein:1977ue}
L.~Wolfenstein,
Phys.\ Rev.\ D \textbf{17}, 2369 (1978);
S.~P.~Mikheev and A.~Y.~Smirnov,
Sov.\ J.\ Nucl.\ Phys.\  \textbf{42}, 913 (1985)
[Yad.\ Fiz.\  \textbf{42}, 1441 (1985)];
Nuovo Cim.\ C \textbf{9}, 17 (1986).

\bibitem{Osland:1999et}
P.~Osland and T.~T.~Wu,
Phys.\ Rev.\ D \textbf{62}, 013008 (2000)
[arXiv:hep-ph/9912540].

\bibitem{Bateman}
Bateman Manuscript Project, \textit{Higher Transcendental Functions},
A. Erd\'elyi, ed.\ (McGraw-Hill, New York, 1953) vol.\ I.

\bibitem{Petcov:1987zj}
S.~T.~Petcov,
Phys.\ Lett.\ B \textbf{200}, 373 (1988).

\bibitem{Bahcall:2000nu}
J.~N.~Bahcall, M.~H.~Pinsonneault and S.~Basu,
Astrophys.\ J.\  \textbf{555}, 990 (2001)
[arXiv:astro-ph/0010346].

\bibitem{Bahcall:1996qv}
J.~N.~Bahcall, E.~Lisi, D.~E.~Alburger, L.~De Braeckeleer, 
S.~J.~Freedman and J.~Napolitano,
Phys.\ Rev.\ C \textbf{54}, 411 (1996)
[arXiv:nucl-th/9601044];
J.~N.~Bahcall,
Phys.\ Rev.\ C \textbf{56}, 3391 (1997)
[arXiv:hep-ph/9710491];
J.~N.~Bahcall and R.~K.~Ulrich,
Rev.\ Mod.\ Phys.\  \textbf{60}, 297 (1988);
J.~N.~Bahcall, M.~Kamionkowski and A.~Sirlin,
Phys.\ Rev.\ D \textbf{51}, 6146 (1995)
[arXiv:astro-ph/9502003];
see also J. N. Bahcall's homepage, 
http://www.sns.ias.edu/$^\sim$jnb.

\bibitem{Fukuda:1998ua}
Y.~Fukuda \textit{et al.}  [Super-Kamiokande Collaboration],
Phys.\ Rev.\ Lett.\  \textbf{82}, 2430 (1999)
[arXiv:hep-ex/9812011];
Phys.\ Rev.\ Lett.\  \textbf{86}, 5656 (2001)
[arXiv:hep-ex/0103033].

\bibitem{Osland:2000gi}
P.~Osland and G.~Vigdel,
Phys.\ Lett.\ B \textbf{488}, 329 (2000)
[arXiv:hep-ph/0006343].

\bibitem{Hata:1993rk}
N.~Hata and P.~Langacker,
Phys.\ Rev.\ D \textbf{48}, 2937 (1993)
[arXiv:hep-ph/9305205];
Phys.\ Rev.\ D \textbf{50}, 632 (1994)
[arXiv:hep-ph/9311214].

\bibitem{Gonzalez-Garcia:2000sq}
M.~C.~Gonzalez-Garcia, M.~Maltoni, C.~Pena-Garay and J.~W.~Valle,
Phys.\ Rev.\ D \textbf{63}, 033005 (2001)
[arXiv:hep-ph/0009350].

\bibitem{Fogli:2001xt}
G.L.~Fogli, E.~Lisi, A.~Marrone, D.~Montanino and A.~Palazzo,
arXiv:hep-ph/0104221.

\bibitem{Apollonio:1997xe}
M.~Apollonio \textit{et al.}  [CHOOZ Collaboration],
Phys.\ Lett.\ B \textbf{420}, 397 (1998)
[arXiv:hep-ex/9711002];
Phys.\ Lett.\ B \textbf{466} (1999) 415
[arXiv:hep-ex/9907037].

\bibitem{Koide:2002nc}
Y.~Koide and H.~Fusaoka,
arXiv:hep-ph/0209148.

\bibitem{Giunti:2002pp}
C.~Giunti and M.~Tanimoto,
arXiv:hep-ph/0209169.

\end{thebibliography}
\end{document}